\documentstyle[epsfig]{aipproc}

\def\1r{P_{33}(1232)}
\def\2r{P_{11}(1440)}
\def\3r{D_{13}(1520)}
\def\4r{S_{11}(1535)}

\begin{document}
\title{Meson photoproduction in the first and second resonance region}

\author{B. Krusche$^*$}
\address{$^*$Department of Physics and Astronomy, University of Basel\\
CH-4056 Basel, Switzerland}

\maketitle

\begin{abstract}

\end{abstract}
The study of baryon resonances via meson photoproduction reactions on the free
proton and on nucleons in the nuclear medium is discussed. Special emphasis is
laid on the production of neutral mesons which due to the suppression of
non-resonant backgrounds are particularly well suited for the study of excited
states of the nucleon. Experiments carried out during the last ten years with
the TAPS-detector at the Mainz Microtron (MAMI) have very significantly
contributed to a detailed investigation of the low lying nucleon
resonances $P_{33}(1232)$, $D_{13}(1520)$ and $S_{11}(1535)$. The most recent
results from single and double pion production and from $\eta$-photoproduction
are summarised.

\section*{Introduction}
Nucleons are complicated many body systems of valence quarks, sea quarks
and gluons interacting via the strong force which in this energy range cannot be
treated in a perturbative manner.
The structure of the nucleon is therefore intimately connected to QCD in the 
nonperturbative range. The study of nucleon resonances plays the same role 
for our understanding of the nucleon structure as nuclear spectroscopy did for
atomic nuclei. In both cases the crucial tests of models come from the
investigation of transitions between the states which are much more sensitive to
model wavefunctions than the excitation energies of the states. From the
experimental point of view the main difference between nuclear and nucleon
structure studies results from the large, overlapping widths of the nucleon
resonances and the much more important non-resonant background contributions
which complicate detailed investigations of individual resonances. In view of
both problems it is very desirable to excite the nucleon resonances via 
different reactions and to study their decays into as many as possible final
states.  Originally most resonances have been identified in pion scattering
experiments wich profit from the large hadronic cross sections. However, this
reaction makes no use of the rich information connected to electromagnetic
transition amplitudes and experimental bias may arise for nucleon resonances
that couple only weakly to th $N\pi$-channel. A comparison of the nucleon 
excitation spectrum predicted by modern, relativistic quark models to the
experimentally established set of nucleon resonances indeed results in the
so-called 'missing resonance' problem: many more states are predicted than
observed. 

The progress made during the last ten years in accelerator and detector
technologies has largely enhanced our possibilites to investigate the nucleon
with different probes. The new generation of electron accelerators CEBAF at
TJNAF in Newport News, ELSA in Bonn, ESRF in Grenoble, MAMI in Mainz and now
also SPring-8 in Osaka, all equipped with tagged photon facilities and 
state-of-the-art detector systems  have opened the way to meson photoproduction
experiments of unprecendented sensitivity and precision. It is interesting to
note that neutral meson photoproduction moved into the center of interest.
Reactions involving neutral mesons have the advantage that non-resonant
background contributions are much less important since the photon couples only
to chargd mesons.

With these new tools two different experimental strategies can be followed.
The problem of missing resonance can be attacked by a large scale survey
investigating many different final states over a large energy range. Such a
program is underway at TJNAF with the CLAS-detector and a complementary program
with the main emphasis on neutral final states with the Crystal Barrel and the
TAPS detectors at ELSA started this year. 

\begin{figure}[h] 
\centerline{\epsfig{file=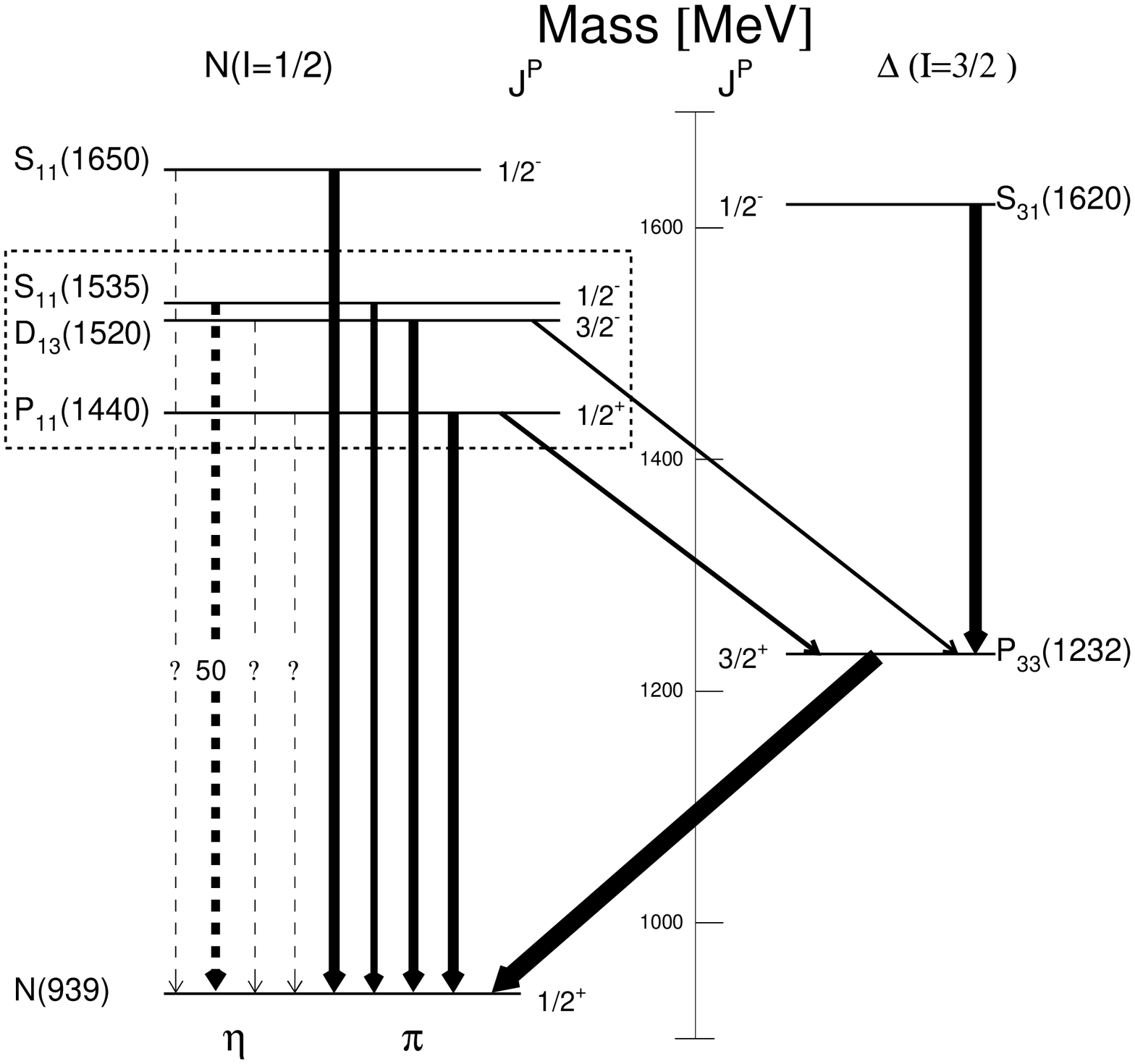,width=3.5in}}
\vspace{10pt}
\caption{Decay scheme of low lying nucleon resonances.
The solid arrows indicate decays via pion emission,
the dashed arrows via $\eta$-emission. The line width of the arrows is scaled
to the branching ratios of the respective decays. 
}
\label{fig1}
\end{figure}
Alternatively the low lying resonances
$\1r$, $\2r$, $\3r$ and $\4r$ can be studied in great detail for precision
tests of the models. The relevant low energy 'level scheme' of the nucleon with
the known decay paths is summarised in figure \ref{fig1}. It is obvious that the
investigation of different meson photoproduction reactions is very advantagous.
One example is the $\4r$-resonance. It's contribution to pion photoproduction is
very small, but it completely dominates $\eta$-photoproduction. The sensitivity
of these experiments has been pushed to limits which just a few years ago were
unimaginable. Let's take as one example the decay branching ratio of the
$\3r$-resonance into the $N\eta$-channel, which as indicated in the figure 
was completely unknown until very recently. This branching ratio must be small 
since the
$\3r$-resonance is located close to the $\eta$-production threshold and the decay
involves an $N\eta$-pair with relative orbital momentum $l=2$. A very precise
study of the angular distributions of the $p(\gamma ,\eta)p$ reaction by the
TAPS group \cite{Krusche_1} revealed for the first time a contribution 
of this resonance to
$\eta$-production. The subsequent measurement of the photon beam asymmetry of
$\eta$-production by the GRAAL group \cite{Ajaka}, which was discussed at this
workshop by E. Hourany, turned out to be so sensitive to $\3r$-contributions that
a branching ratio of 0.05\% - 0.08\% could be extracted \cite{Tiator}.

Much less is yet known about the behavior of the nucleon resonances inside the
nuclear medium.  Modifications can arise from a variety of different aspects.
The most trivial effect is the broadening of the excitation functions due to 
nuclear Fermi motion. The decay of the resonances is modified by Pauli-blocking
of final states, which reduces the resonance widths, and by additional decay
channels like $N^{\star}N\rightarrow NN$ which cause the so-called collisional
broadening. Both effects cancel to some extend and it is a priori not clear
which one will dominate.
A very exciting perspective is that the resonance widths can be sensitive 
to in-medium mass modifications of mesons arising from chiral restoration 
effects. The $D_{13}$-resonance e.g. has a non negligible
decay branching ratio into the $N\rho$-channel \cite{PDG}.
Very recent results from the $p(\gamma, \pi^o\pi^+)n$ reaction 
\cite{Langgaertner} even suggest that this contribution might be larger 
than expected. 
This means that a broadening or a downward shift of the $\rho$-mass distribution
inside the nuclear medium could have significant effects on the $D_{13}$ width.

The first experimental investigation of the second resonance region for nuclei
was done with total photoabsorption. The surprising results showed
an almost complete depletion of the resonance bump \cite{Frommhold,Bianchi}
which up to now has not been fully understood. 
Total photoabsorption has the advantage, that no final state
interaction effects (FSI) must be accounted for so that the entire nuclear
volume is tested. However, many different reaction channels do contribute to
this reaction so that it is impossible to test the behavior of individual
resonances. Even worse, as discussed below some of the reaction channels 
are not strongly related to the excitation of resonances from the second 
resonance region. 
It is therefore desirable to study this region with exclusive reactions which
allow the investigation of individual resonances, even at the expense that FSI
effects complicate the interpretation.

In the present talk I will mainly concentrate on a few typical examples for the
investigation of nucleon resonances with meson photoproduction reactions on the
free nucleon and on nucleons bound in nuclei which were carried out during the
last few years at the Mainz MAMI accelerator.

\section*{Experiments}
The experiments discussed here were mostly carried out at the Glasgow 
tagged photon
facility installed at the Mainz microton MAMI. The tagged photon facility 
uses Bremstrahlung photons produced with the 850 MeV electron beam in a
radiator foil. The standard tagging range covers photon energies between 
50 and 790 MeV, although for many experiments the low energy section of the
tagger is switched off, to allow for higher intensities at high photon energies.
This is possible since usually the electron beam intensity is limited by the 
fastest counting photomultipliers in the tagger focal plane at intensities
roughly two orders of magnitude below the capabilities of the electron machine.
The maxium tagged photon energies achieved so far are 820 MeV 
(at an electron beam energy of 880 MeV), however this will be increased to
approximately 1.4 - 1.5 GeV after the machine upgrade starting in spring 2001.

\begin{figure}[h] 
\centerline{\epsfig{file=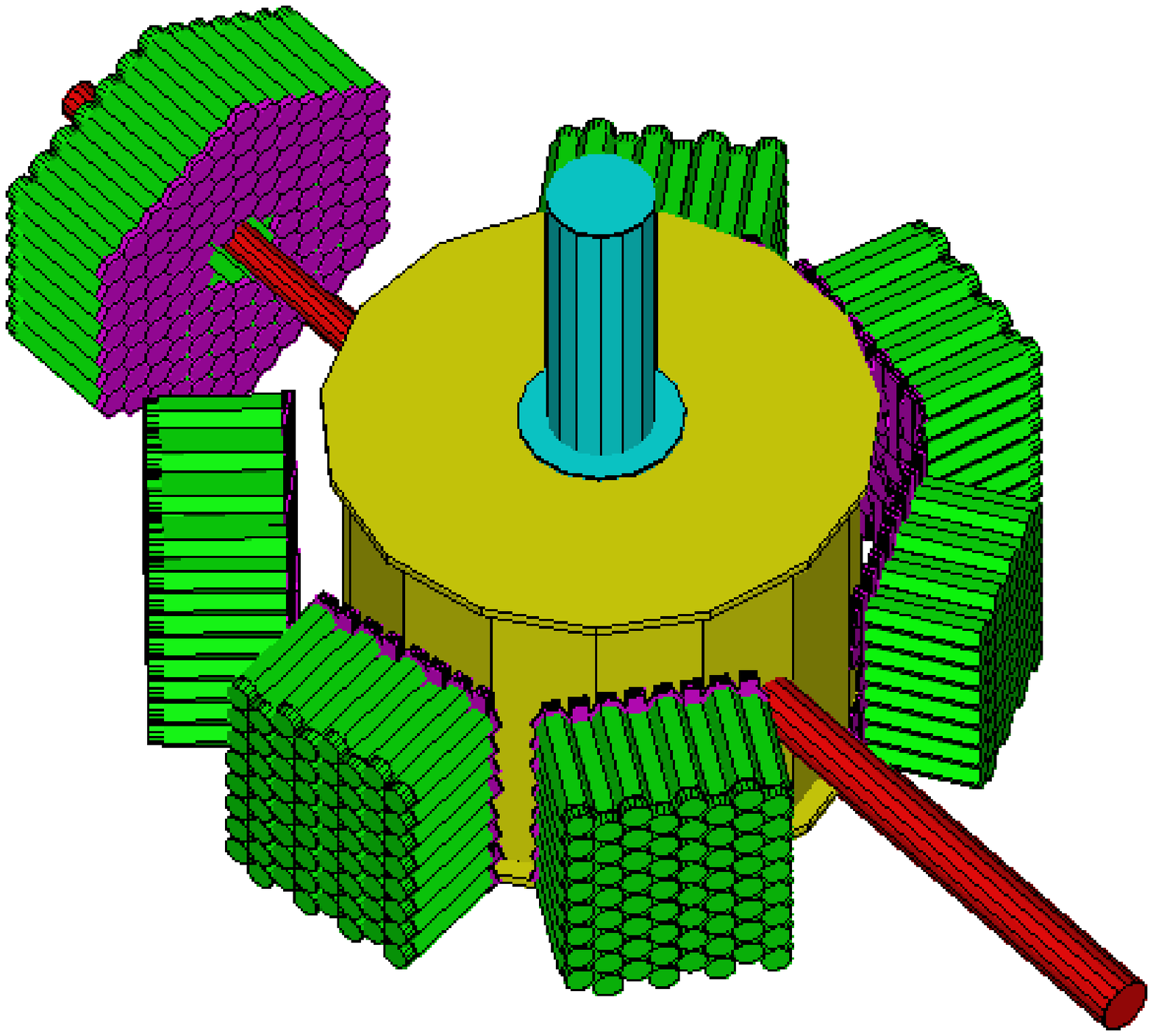,width=4.6in}}
\caption{Arrangement of the TAPS-detector at the MAMI accelerator for the
experiments carried out in 1995/1996. Six blocks each consisting of 64 BaF$_2$
modules and a forward wall with 120 BaF$_2$-modules were arranged in one plane
around the scattering chamber.
}
\label{fig2}
\end{figure}

The typical focal plane energy resolution is limited by the geometrical width
of the focal plane counters to approximately 2 MeV. 
However, since the intrinsic resolution of
the magnet is much better (roughly 100 keV), for some threshold experiments
'tagger microscopes' with scintillation counters of much smaller width have
been used to improve the energy resolution to a few hundred keV. Circular
polarized photon beams are routinely available over the full energy range,
linearly polarized photon beams produced via coherent Bremsstrahlung from
diamonds are available in the $\Delta$-resonance region.

Most of the meson production experiments presented in this talk were carried
out with
the photon detector TAPS. The detector consists of more than 500 hexagonally
shaped BaF$_2$ scintillators of 25 cm length corresponding to 12 radiation 
lengths. The scintillators are equipped with individual plastic veta detectors 
for the discrimination of charged particles.
One of the possible configurations, which was used in Mainz, is shown in fig.
\ref{fig2}. In this configuration six block structures each consisting of 64
crystals in an $8\times 8$ matrix were combined with a forward wall of 120
crystals. The device is optimized for the detection of photons via
electromagnetic showers, but has also particle detection capabilites. The
separation of photons from massive particles makes use of the plastic veto 
detectors (only charged particles), a time-of-flight measurement with typically
500 ps resolution (FWHM) and the excellent pulse shape discrimination
capabilities of BaF$_2$-scintillators. The combination of these methods produces
extremly clean samples of the meson decay photons. The identification of neutral
mesons ($\pi^o$ and $\eta$) then makes use of a standard invariant mass 
analysis.

\section*{Results and Discussion}

\subsection*{The $\Delta$-resonance region}
The $\1r$-isobar is doubtlessly the best studied excited state of the nucleon. 
Is there anything more to learn about it with more sensitive or precise
experiments? An example for recent progress is the measurement of the
$E2$-admixture in its excitation.  
It is well known that in photoproduction reactions the $I=J=3/2$ 
$\Delta$-excitation of the nucleon is dominated by the magnetic 
multipole $M_{1+}$. In the simplest picture the incident $M1$ photon 
induces the spin-flip of one of the constituent quarks. However, as far as 
quantum numbers are concerned the excitation via an $E2$-photon
($E_{1+}$-multipole) is also possible. An $E2$-admixture in the transition
strength could e.g. arise from d-state components in the baryon wave
functions connected to tensor forces. The correct prediction of such admixtures
is a challenge for nucleon models. 

Experimentally one must determine the ratio $R_{EM}=E_{1+}^{3/2}/M_{1+}^{3/2}$
of the electric quadrupole to the magnetic dipole in the isopin 3/2 channel.
This may be accomplished by a measurement of the differential cross sections and
the photon beam asymmetries in the $p(\vec{\gamma} ,\pi^o )p$ and the
$p(\vec{\gamma} ,\pi^+ )n$ reactions. Such measurements were performed at MAMI
\cite{Beck97} with the DAPHNE detector and at LEGS \cite{Blanpied97}.  
Although the extraction of the ratio is not completly model independent 
(see e.g. discussion in \cite{Beck00} and refs. therein) all analyses find a
small negative value of the $R_{EM}$-ratio. The results from the Mainz data
$-(2.5\pm 0.1(stat)\pm 0.2(sys))\%$ and the LEGS data 
$-(3.1\pm 0.3(stat+sys)\pm 0.2 (model))\%$ are in reasonable agreement.

Here I will discuss two other recent experiments concerning properties of the 
$\Delta$-isobar excited on the free proton and on nucleons bound in nuclei.

\subsubsection*{The reaction $p(\gamma ,\pi^o\gamma)p$ and the magnetic moment
of the $\Delta^+$}
The magnetic moment of the $\Delta$-resonance is one of its properties which are
sensitive to the spin-flavor correlations of the quarks which are related to the
configuration mixing predicted by QCD. In particular in case of the $\Delta$
most models assume a quark structure similar to the nucleon ground state but
with spin and isospin of the quarks coupled to 3/2 instead of 1/2. 
Predictions for the magnetic moment have been made by many models but so far an
experimental value is only available for the $\Delta^{++}$ isobar measured via the
$\pi^+ p\rightarrow\pi^+\gamma p$ reaction \cite{Nefkens,Bosshard}. The most
recent experimental value \cite{Bosshard} of 
$\mu_{\Delta^{++}}=(4.52\pm0.50)\mu_N$
is significantly smaller than the naive constituent quark model prediction of 
$\mu_{\Delta^{++}}=2\mu_p=5.58\mu_N$. The magnetic moment of the $\Delta^+$
isobar was not measured up to now, but it was recently pointed out
\cite{Machavariani,Drechsel0} that the $p(\gamma ,\pi^o\gamma')p$-reaction is particularly
well suited for its measurement.

\begin{figure}[h] 
\centerline{\epsfig{file=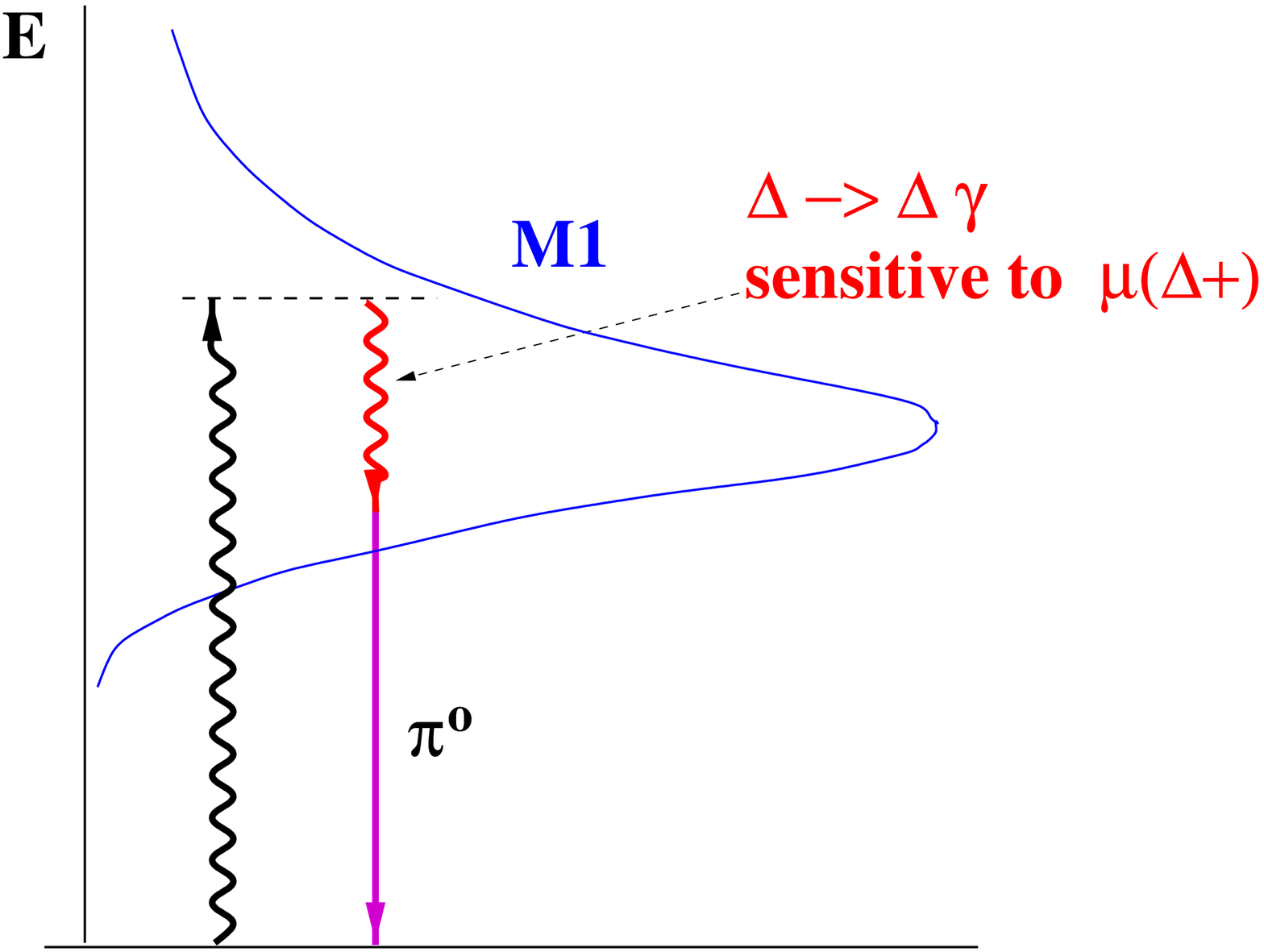,width=4.3in}}
\vspace{10pt}
\caption{Principle of the determination of the $\Delta$ magnetic moment
from the reaction $p(\gamma ,\pi^o\gamma ')p$.
The $\Delta$-resonance is excited by the incident photon at its high energy
tail, de-excites via a M1 reorientation transition and decays to the nucleon
ground state via pion emission. 
}
\label{fig3}
\end{figure}

The principle of this experiment is illustrated in fig. \ref{fig3}.
The $\Delta$-resonance is excited by a real photon, decays within its final
width via an electromagnetic M1 re-alignment transition, which is sensitive to
the magnetic moment, and finally 
de-excites by emission of a $\pi^o$-meson to the nucleon ground state.
There are of course also background diagrams which contribute to the   
$p(\gamma ,\pi^o\gamma)p$-reaction, but it is pointed out in 
\cite{Machavariani}
that the angular distributions of this reaction are sufficently sensitive
to the magnetic moment. 

The experimental identification of this reaction is not simple. In the
energy range of interest background from two different sources must be
eliminated.
Background originates from double $\pi^o$-photoproduction events where one 
photon has
escaped due to the finite solid angle coverage. Further background is caused
by events from single $\pi^o$-photoproduction, where the electromagnetic shower
from one photon has split off a satelite which is missidentified as independent
photon. Such split-offs are not abundent, but the cross section for single
$\pi^o$-photoproduction ($\approx 300 \mu b$ at maximum) is larger by
roughly four orders of magnitude than the reaction of interest 
(roughly 10nb \cite{Machavariani,Drechsel0}).
Nevertheless a very clean separation of the reaction was recently achieved
\cite{Kotulla}. In the first step veto detectors, time-of-flight, time-of-flight
versus energy and the pulseshape analysis were exploited to identify events 
with
exactly three photons and one proton detected. The $\pi^o$-meson was identified
via an invariant mass analysis. 
\begin{figure}[h] 
\begin{minipage}{0.cm}
{\epsfig{file=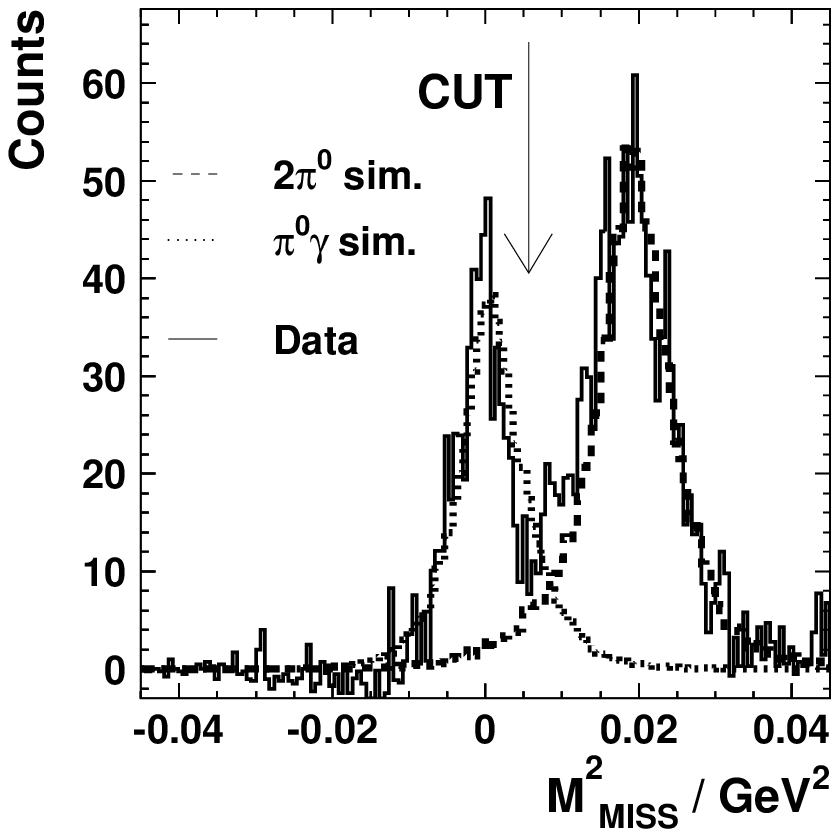,height=2.9in,width=2.9in}}
\end{minipage}
\hspace{7.2cm}
\begin{minipage}{0.cm}
{\epsfig{file=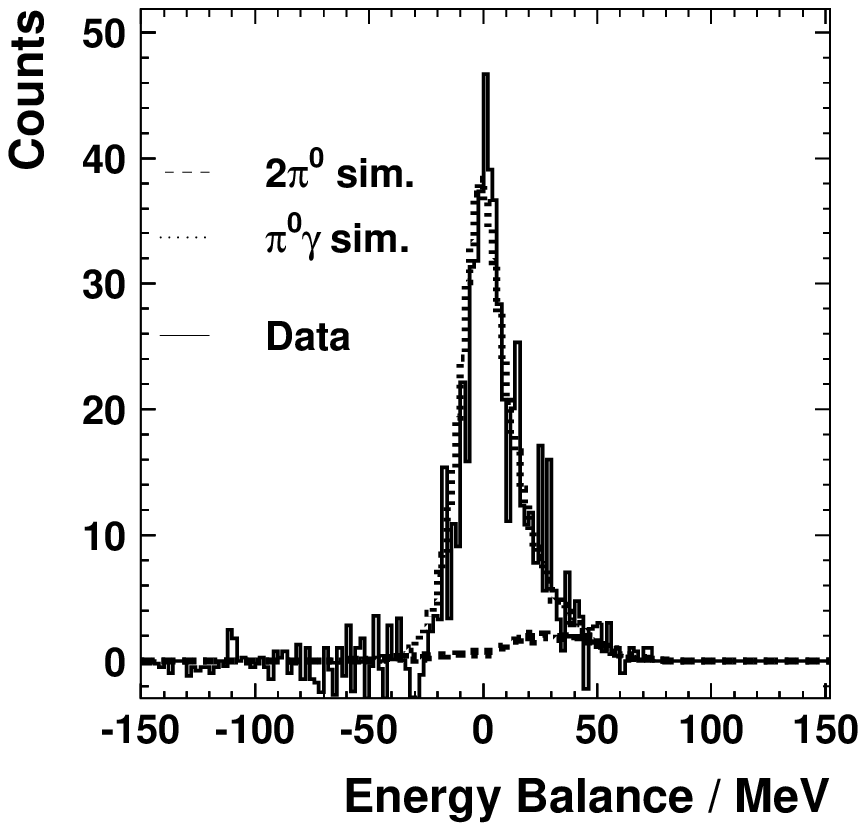,height=2.9in,width=2.9in}}
\end{minipage}
\vspace{10pt}
\caption{Identification of the reaction $p(\gamma ,\pi^o\gamma ')p$ for incident
photon energies in the range 245-475 MeV. The left
hand side shows a missing mass spectrum calculated from the energy of the
incident photon and the four-momenta of the detected $\pi^o$-meson and the
proton. 'Good' events have zero missing mass, background from double
$\pi^o$-production appears at the pion mass. The right hand side shows the
the energy balance spectrum (difference of initial and final state energy)
for those events that passed the cut in the msissing mass
spectrum. In both pictures the dotted [dashed] lines indicate Monte Carlo
simulations of the  $p(\gamma ,\pi^o\gamma ')p$ [$p(\gamma ,2\pi^o)p$]
reactions.
}
\label{fig4}
\end{figure}
Subsequently, the missing mass was constructed
from the energy of the incident photon and the four-momentum of the proton.  
Events with missing mass equal to the pion mass were eliminated since they
are due to shower split-offs. It was required in the next step that the sum of
the momenta of the three photons and the proton vanished in x- and y-direction
and was equal to the momentum of the incident photon in z- (beam) direction, all
of course within the experimental resolution.
The two final steps of the identification are shown in fig. \ref{fig4}.
The missing mass was calculated 
from the energy of the incident photon and the four-momenta of the pion and the
proton and it was made use of energy conservation by a comparison of the
total initial and final state energies. It is evident from the figure that this
procedure produces an almost background free data sample for the   
$p(\gamma ,\pi^o\gamma ')p$-reaction. The analysis of the reaction in view of
the $\Delta^+$ magnetic moment is under way.

\subsubsection*{Coherent $\pi^o$-photoproduction from heavy nuclei - the 
$\Delta$ in the nuclear medium}
The study of meson photoproduction from atomic nuclei is mainly 
motivated by two strongly interconnected aspects namely possible 
medium modifications of the excited states of the nucleon and the 
meson-nucleus interaction. Particularly interesting is the case where 
the reaction amplitudes from all nucleons add up coherently and the 
nucleus remains in its ground state after the reaction. The 
theoretical treatment of this process involves much fewer assumptions 
and approximations than are needed for the description of complicated 
final states arising e.g. from breakup reactions where the 
participant nucleon is knocked out of the nucleus.

\begin{figure}[h] 
\begin{minipage}{0.cm}
{\epsfig{file=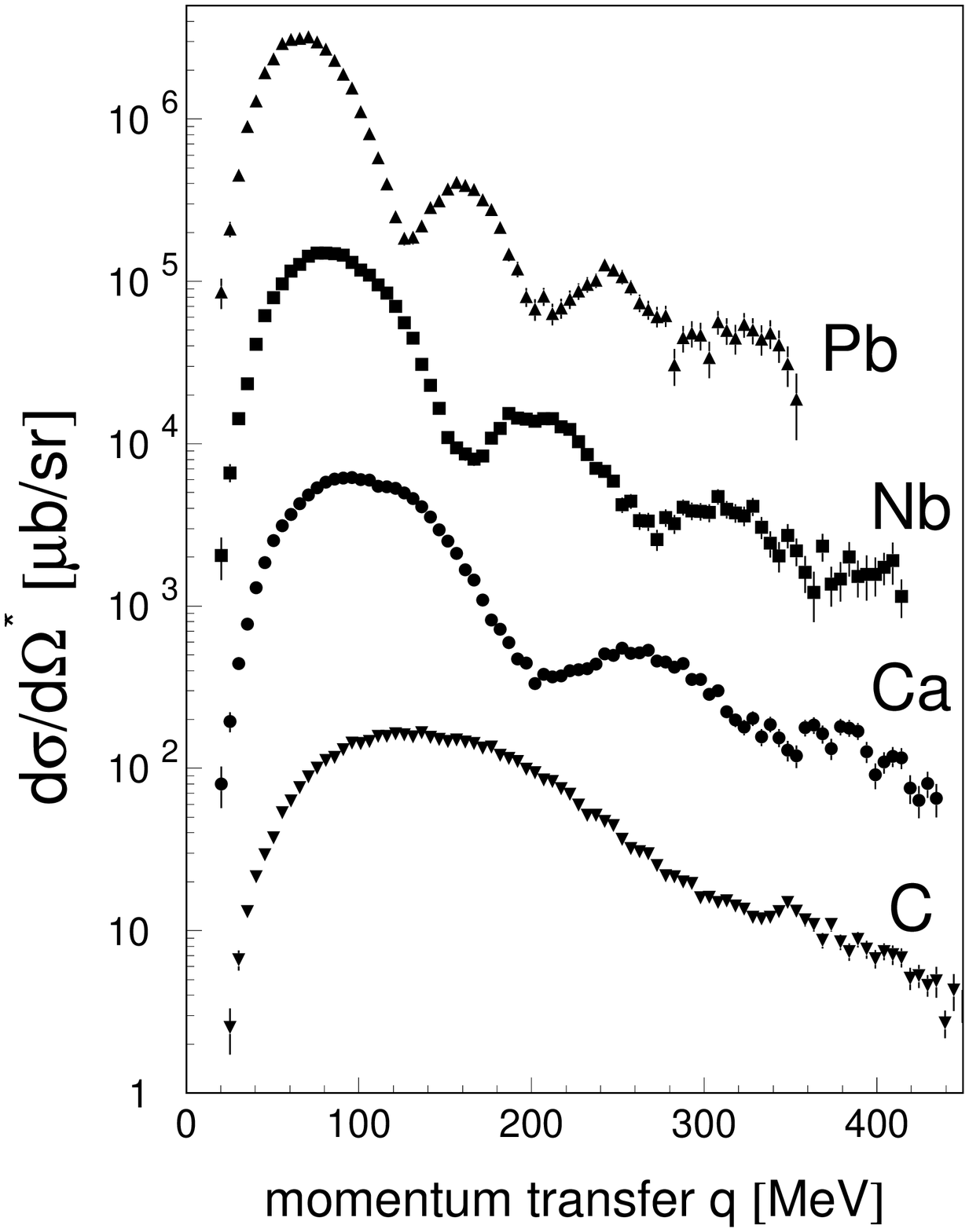,height=3.4in,width=2.6in}}
\end{minipage}
\hspace{6.3cm}
\begin{minipage}{0.cm}
{\epsfig{file=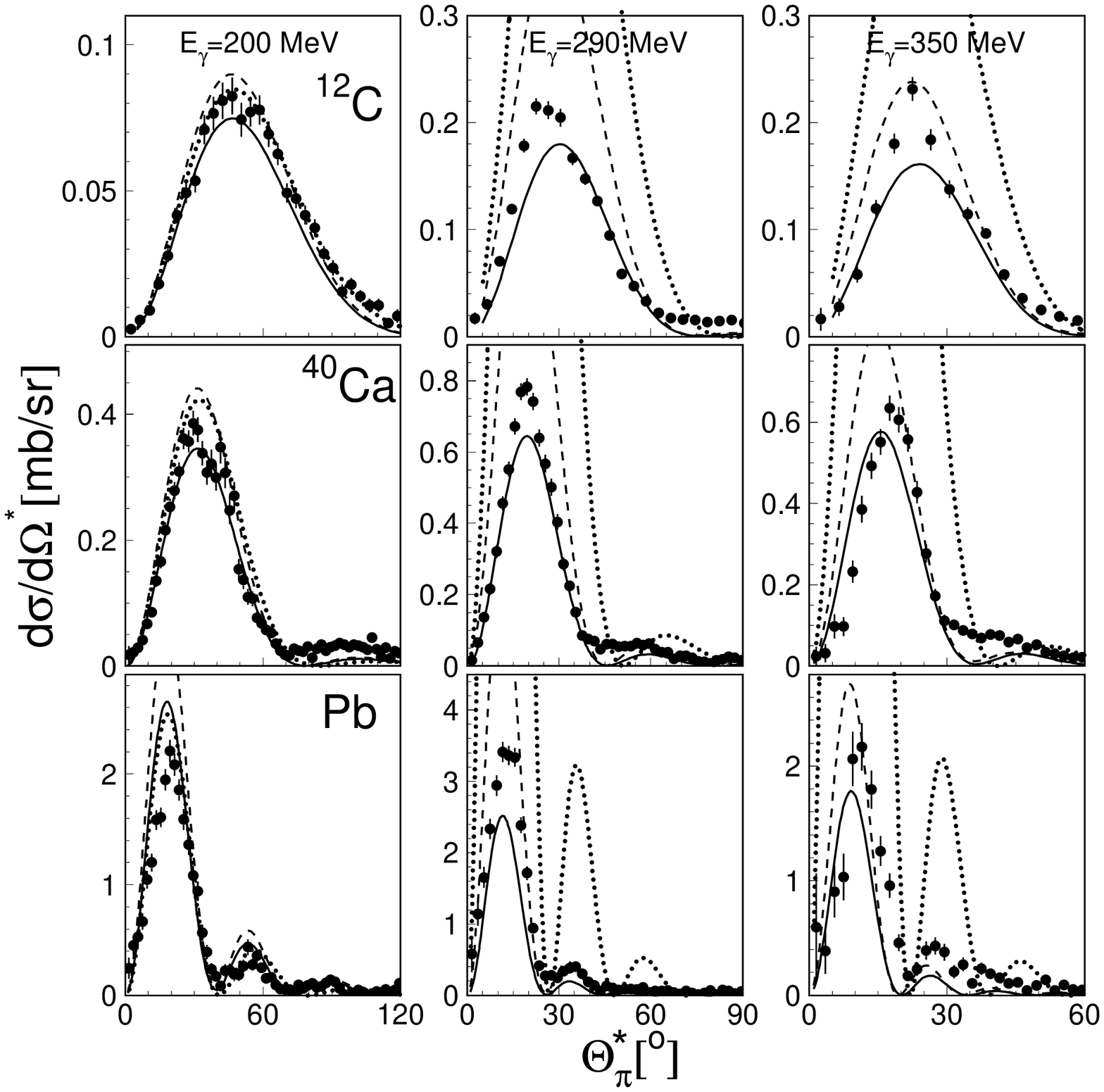,height=3.4in,width=3.2in}}
\end{minipage}
\vspace{10pt}
\caption{Coherent $\pi^o$-photoproduction from heavy nuclei in the
$\Delta$-resonance region. Left hand side: differential cross
sections as function of momentum transfer averaged from 200 - 290 MeV incident
photon enertgies. Right hand side: angular distributions for $^{12}C$, 
$^{40}Ca$ und $Pb$ compared to model predictions \protect\cite{Drechsel}.
Dotted line: PWIA, dashed: DWIA, full lines: DWIA with $\Delta$-self energy
fitted to $^4$He cross sections. 
}
\label{fig5}
\end{figure}
In a recent experiment with TAPS at MAMI coherent 
$\pi^{o}$-photoproduction from $^{12}C$, $^{40}Ca$, $^{93}Nb$ and 
$^{nat}Pb$ was studied throughout the $\Delta$-resonance region 
\cite{Krusche_2}. 

The characteristic features of the coherent process
from spin 0 nuclei
in the most simple PWIA approximation are the proportionalities to the 
nuclear mass form factor, to the square of the atomic mass number and 
to  $sin^{2}(\Theta_{\pi})$. The form factor dependence and the 
$sin^{2}$-term, which forces the forward cross section to zero
are clearly visible in the left hand side of fig. \ref{fig5} where 
the differential cross sections averaged over photon energies from 
200 - 290 MeV are plotted versus the momentum transfered to the 
nucleus. The $A^{2}$-dependence was demonstrated at incident photon 
energies around 220 MeV by a comparison to the PWIA prediction 
\cite{Krusche_2}, it is 
obscured at higher incident photon energies by the pion-nucleus final 
state interaction.

A detailed investigation of the FSI effects and possible medium modifications of
the $\Delta$-resonance requires an analysis far beyond PWIA. Recently Drechsel
and coworkers \cite{Drechsel} developed a model for coherent
$\pi^o$-photoproduction starting from their Unitary Isobar Model
for the elementary reaction \cite{Drechsel2}. Final state interactions of the
pions are taken into account via a distorted wave impulse approximation (DWIA)
and in addition medium modifications of the $\Delta$ are included via a
phenomenological parametrization of the $\Delta$ self-energy. The potential used
for the self-energy was fitted to differential cross sections of the
$^4He(\gamma ,\pi^o)^4He$ reaction \cite{Rambo} and then kept fixed for the
prediction of cross sections for heavy nuclei. The results are compared to the
data in the right hand side of Fig. \ref{fig5}.
At low incident photon energies ($E_{\gamma}=200$ MeV) the difference between 
PWIA, DWIA and additional $\Delta$-modification are small. However, the 
cross sections are strongly overestimated around the $\Delta$-resonance 
position by the PWIA and DWIA calculations. Reasonable agreement is only 
achieved when the $\Delta$-nucleus interaction is taken into account. 
The $\Delta$-self energy extracted from the $^4He$ data for this incident 
photon energy ($E_{\gamma}=290$ MeV)is $Re(V)\approx 19$ MeV and 
$Im(V)\approx -33$ MeV \cite{Drechsel}, corresponding to a significant effective
broadening of the resonance by 66 MeV.
Based on a comparison of their prediction to the few data then available 
for $^{12}C(\gamma ,\pi^o )^{12}C$ it was suggested by Drechsel et al. 
\cite{Drechsel} that the $\Delta$-nucleus interaction already saturates for 
$^4He$. The present data demonstrate that indeed the A-dependence of the 
potential is not large since the agreement between model predictions and data 
is comparable from carbon up to lead although all data are somewhat 
underestimated by the calculations. Finally at energies above the 
$\Delta$-resonance position ($E_{\gamma}=350$ MeV) the sensitivity to
$\Delta$-modifications is small and the full model is not in significantly
better agreement with the data than the DWIA calculation.  

\subsection*{The second resonance region}
The so-called second resonance region of the nucleon includes the
$P_{11}(1440)$-, the $D_{13}(1520)$-, and the $S_{11}(1535)$-resonances which
can be excited e.g. by photons in the energy range 600 - 900 MeV. Due to
their close spacing and their relatively large widths (100 - 300 MeV)
all resonances overlap.  However, due to their
different couplings to the initial $\gamma N$ state and to the final 
meson-nucleon states, different resonances are dominating the possible 
meson photoproduction reactions. The production of $\eta$-mesons
proceeds almost exclusively via the excitation of the $S_{11}(1535)$ resonance,
while the largest resonance contributions to single and double pion 
production come from the $D_{13}(1520)$ resonance. Using this selectivity
the properties of these two resonances, when excited on the free proton
or quasifree neutron, have been studied in much detail during the last 
few years via $\eta$-photoproduction 
\cite{Krusche_1,Krusche_2,Bock,Ajaka,Krusche_3,Hoffmann,Hejny} 
and single and double pion photoproduction reactions 
\cite{Braghieri,Tejedor,Haerter,Zabrodin,Krusche_4,Kleber,Wolf,Langgaertner}.

\subsubsection*{$\eta$-photoproduction from the proton and the $S_{11}(1535)$}

As already mentioned $\eta$-photoproduction is the method of choice to
investigate the $S_{11}(1535)$ resonance. 
\begin{figure}[h] 
\begin{minipage}{0.cm}
{\epsfig{file=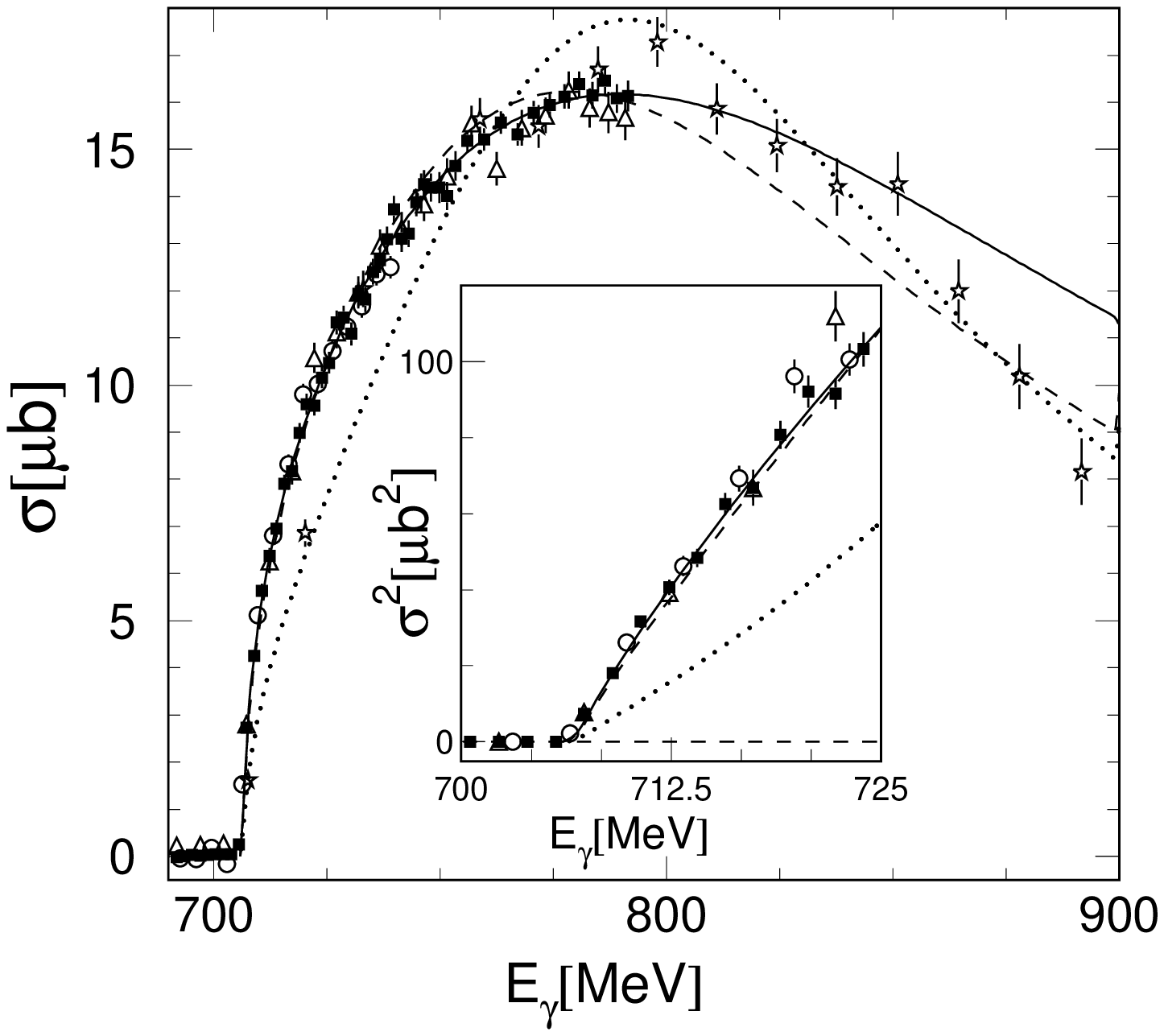,height=3.0in,width=2.9in}}
\end{minipage}
\hspace{7.2cm}
\begin{minipage}{0.cm}
{\epsfig{file=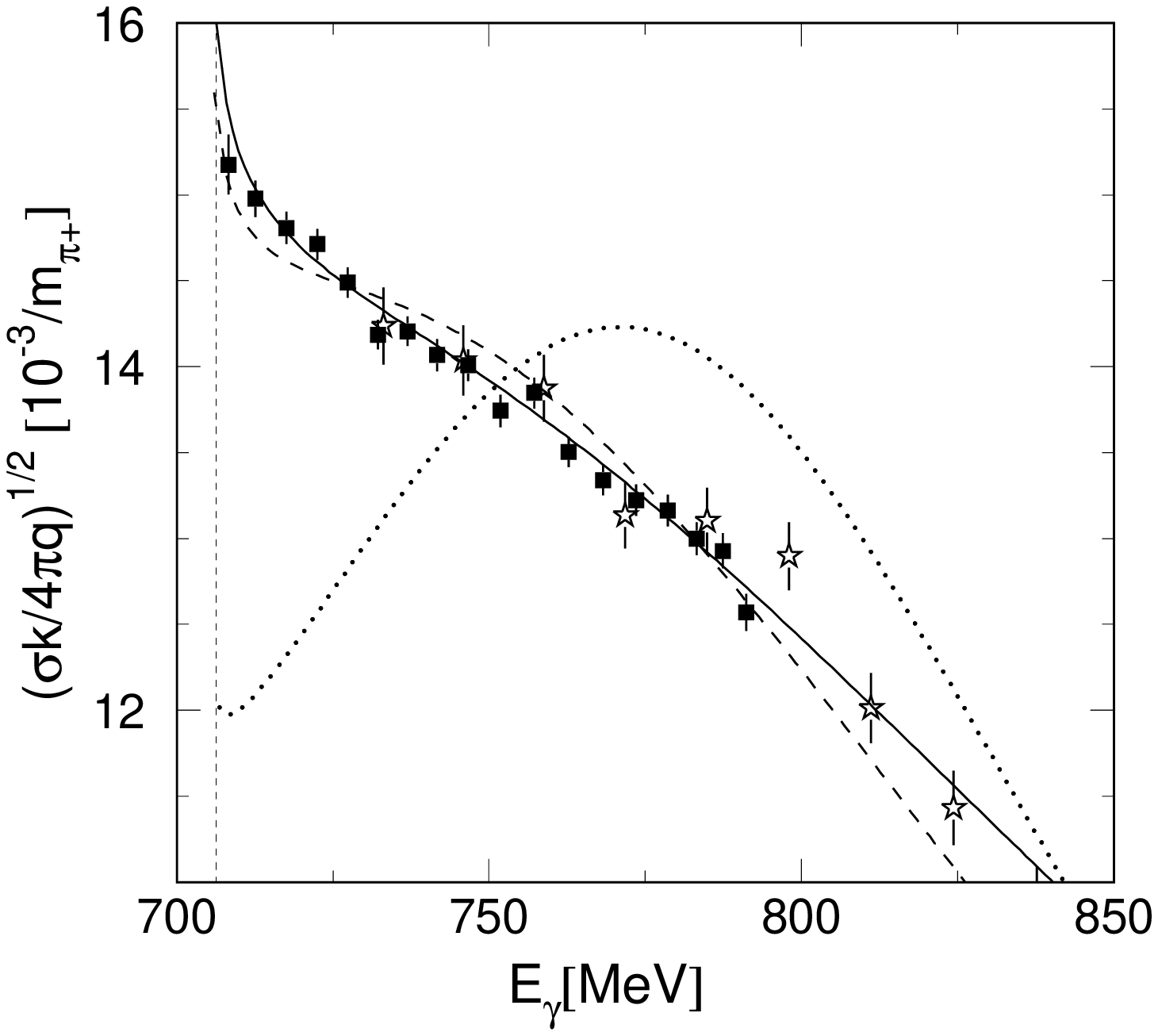,height=3.0in,width=2.9in}}
\end{minipage}
\vspace{10pt}
\caption{Breit-Wigner fits of total cross section of $\eta$ photoprodcution from
the proton. The left hand side shows the total cross section. The data are from
\protect\cite{Wilhelm} (stars) and \protect\cite{Krusche_1} (all other). 
The curves correspond to Breit-Wigner curves with widths of 200 MeV 
(full curve), 151 MeV (dashed) and 112 MeV (dotted). The insert shows the
squared threshold cross section which is expected to depend linearly on the
photon energy. The right hand side shows the square root of the phase space
reduced cross section which is proportinal to $|E_{o+}|$ as long as the $S_{11}$
excitation dominates the cross section. Data and curves are the same as on the
left hand side.
}
\label{fig6}
\end{figure}
 Recently there was some discussion
about the parameters of this resonance in particular its electromagnetic 
helicity amplitude $A^p_{1/2}$ and the width $\Gamma_{S_{11}}$ which
is also of importance for the interpretation of the nuclear data as reported by
J. Kasagi at this workshop. The range given by the particle data group for the
width as derived from pion production experiments is 100 - 250 MeV. The
Breit-Wigner fit of the Mainz and Bonn data shown in Fig. \ref{fig6} (solid line)
gave a
width close to 200 MeV \cite{Krusche_1,Krusche_2}. This fit describes the 
data very well up to photon 
energies of 850 MeV, but it overestimates the data at higher energies.
Here one must keep in mind that as pointed out in \cite{Sauermann} at 
the higher energies significant effects are expected from the $S_{11}(1620)$ 
resonance  which are of course not accounted for by the BW-fit. Nevertheless,
due to the overestimation of the free proton data at energies above 850 MeV,
this BW-curve is not well suited for the interpretation of the nuclear data 
over a larger energy region \cite{Yorita}. Analyses of $\pi N$-data have 
sometimes lead to extremely small values of the width.
An example is the multichannel unitary analysis by Dytman et al. \cite{Dytmann} 
which gave a width of 112(30) MeV. A Breit-Wigner curve corresponding to the
$S_{11}$-parameters found in \cite{Dytmann} is shown in the figure as dotted
curve. Obviously this resonance paremeters would require very substantial
non-$S_{11}$ contributions to $\eta$-photoproduction already close to threshold,
which have not been found in other analyses (see e.g. \cite{Krusche_2}). 
Recently, the GRAAL group has measured $\eta$-photoproduction up to higher
energies (see contribution of E. Hournany to this workshop). They found a
reasonable description of the total cross section from the threshold region
up to high energies with a fit corresponding to a resonance position at
1536.8 MeV and a width of 151 MeV. The corresponding BW-curve is shown in Fig.
\ref{fig6} as dashed line. Although the agreement with the threshold data is
slightly worse than for the fit with the larger width, such a resonance curve is
certainly the best compromise for the interpretation of the nuclear data. 

\subsubsection*{$\eta$-photoproduction from light nuclei and the $S_{11}(1535)$
excitation on the neutron}

The photoproduction of $\eta$-mesons from light nuclei ($^2H$,$^4He$,$^3He$) 
was studied for the extraction of the isospin composition of the electromagnetic
$S_{11}$ excitation. The quasifree reaction was used to determine the cross
section from the neutron and thus to extract the ratio of the electromagnetic
helicity amplitudes $A^n_{1/2}/A^p_{1/2}$ for which predictions are available
from quark models. The experimental determination of the ratio is much less
prone to systematic errors than the helicity amplitudes themselves since the
large uncertainites due to the width and the decay branching ratios of the
resonance cancel in the ratio. Quasifree $\eta$-photoproduction was studied in 
inclusive measurements, where the neutron cross section is extracted via a
comparison of the inclusive nuclear cross section and the Fermi smeared proton
cross section, and in exclusive experiments with detection of the recoil protons
and neutrons. The results of these experiments, carried out in Mainz and Bonn,
which are summerised below are in excellent agreement, so that this ratio is
certainly one of the best determined parameters of the $S_{11}(1535)$ resonance: 
\begin{itemize}
\item[*]{$A_{1/2}^{n}/A_{1/2}^{p}=\pm\sqrt{0.66\pm 0.07}$,
$\pm\sqrt{0.68\pm0.07}$  
  ~~~~~~~~(deuterium, inclusive)~   \cite{Krusche_3,Weiss}}
\item[*]{$A_{1/2}^{n}/A_{1/2}^{p}=\pm\sqrt{0.68\pm 0.06}$,
$\pm\sqrt{0.6-0.75}$  
  ~~~~~~~~(deuterium exclusive)~    \cite{Hoffmann,Weiss}}
\item[*]{$A_{1/2}^{n}/A_{1/2}^{p}=\pm\sqrt{0.67\pm (0.01)_{stat}}$ 
  ~~~~~~~~~~~~~~~~~~~~(helium, inclusive)~~~~~      \cite{Hejny} } 
\item[*]{$A_{1/2}^{n}/A_{1/2}^{p}=\pm\sqrt{0.68\pm (0.02)_{stat} \pm (0.09)_{sys}}$ 
  ~~~~~(helium, exclusive)~~~~~      \cite{Hejny}. }
\end{itemize}

Coherent $\eta$-production from the deuteron can be used to extract the
isoscalar part of the transition amplitude. The upper limit for this cross
section obtained in \cite{Krusche_3} and the values reported in
\cite{Hoffmann,Weiss} are so small that they exclude a dominant isoscalar part
so that $A^n_{1/2}/A^p_{1/2}<0$ which fixes the sign of the ratio.
Consequently, $\eta$-photoproduction in the $S_{11}$ region is dominated by an
isovector, spinflip amplitude. This implies that the cross section
for the coherent process from $^3He$ (I=J=1/2) should be considerably larger
than for $^4He$ (I=J=0). The coherent cross section from $^4He$ \cite{Hejny}
is indeed so small that up to now only upper limits could be determined.
However, very preliminary results from a recent measurement of 
$^3He(\gamma ,\eta)X$ with TAPS at MAMI show a clear signal for coherent 
$\eta$-production.  

\subsubsection*{Double pion production and the $D_{13}(1520)$} 
Double pion photoproduction is a very important reaction channel in the second
resonance region. The cross sections for single meson production 
(pions and $\eta$-mesons) and double pion photoproduction are almost equal at
incident photon energies between 600 and 800 MeV. Moreover, most of the rise of
the total photoabsorption cross section from the dip above the
$\Delta$-resonance to the peak of the second resonance bump is due to double
pion production. 
\begin{figure}[h] 
\begin{minipage}{0.cm}
{\epsfig{file=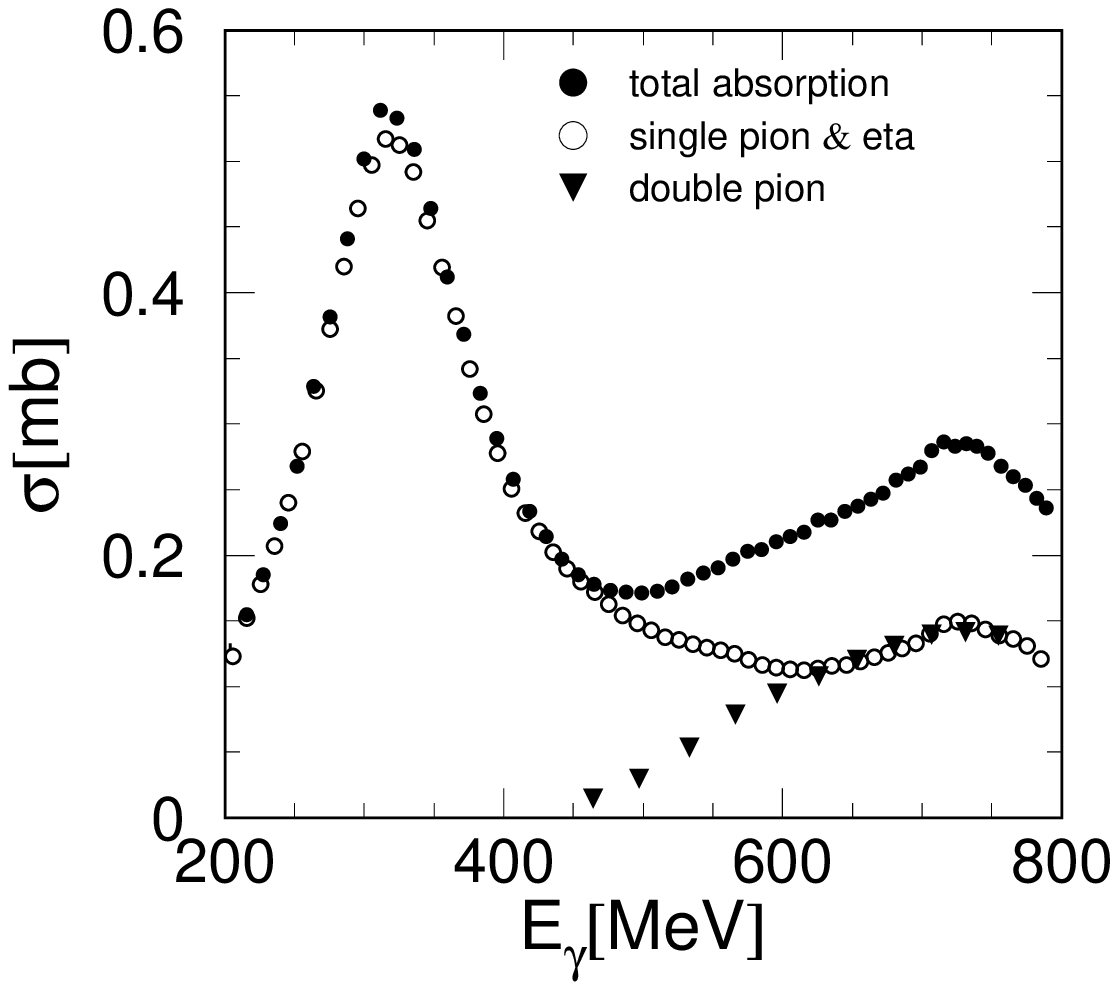,height=2.8in,width=2.8in}}
\end{minipage}
\hspace{7.cm}
\begin{minipage}{0.cm}
{\epsfig{file=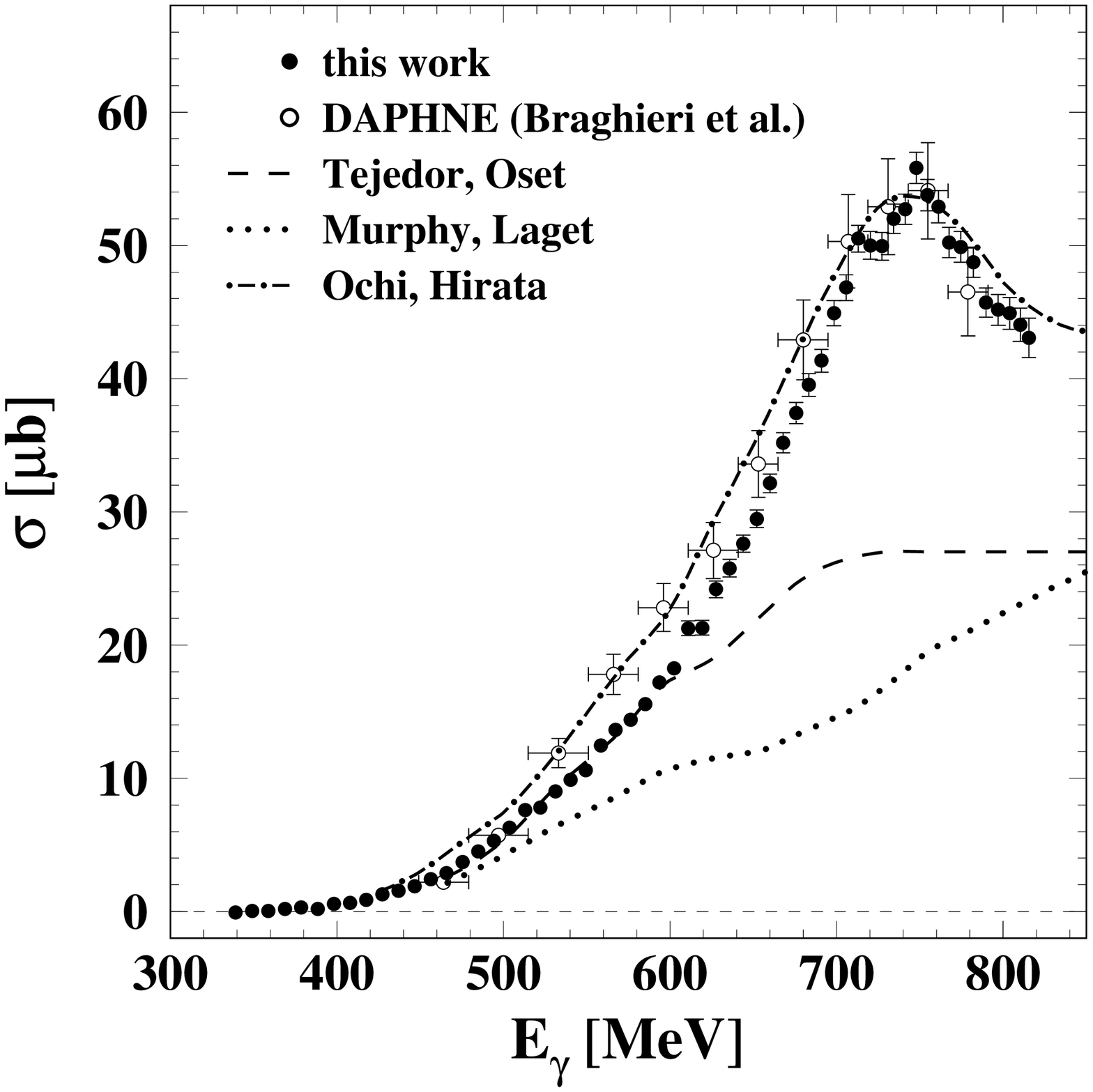,width=2.8in}}
\end{minipage}
\vspace{10pt}
\caption{Left hand side: decomposition of the total photoabsorption cross
section of the proton into single and double meson production reactions.
Right hand side: total cross section of the reaction 
$p(\gamma ,\pi^+\pi^o)n$. The data are from \protect\cite{Braghieri} (DAPHNE) 
and \protect\cite{Langgaertner} (TAPS).
The dashed, dotted and dash-dotted curves are the results of the model
calculations from \protect\cite{Tejedor,Murphy,Ochi}. 
}
\label{fig7}
\end{figure}
This is demonstrated in the left hand side of Fig. \ref{fig7}
where the total photoabsorption cross section of the proton is compared to the 
single and double meson production cross sections. 
Any detailed interpretation of the second resonance bump requires the
understanding of double pion production. This is not only important for
resonances on the free nucleon, but also for the understanding of the
suppression of the second resonance bump in total photoabsorption from nuclei. 
Some authors 
\cite{Giannini} interpreted this as evidence for a damping of the 
excitation of the $P_{11}$, $D_{13}$ and $S_{11}$ in the nuclear medium, others
argued \cite{Lehr} that a strong broadening e.g. of the $D_{13}$-resonance
might result from the coupling to the $N\rho$-channel if the $\rho$-mass
distribution is modified inside the nuclear medium.
However, it was not even clear if these resonances play an important role
for double pion production which makes such a large contribution to the `bump'.
Background terms like the $\Delta$-Kroll-Rudermann (KR) and the 
$\Delta$-pion-pole term which instead involve the excitation of the $\Delta$ 
are important at least for the charged double pion channels.

Among the possible double pion production reactions 
$\gamma p\rightarrow\pi^+\pi^-$is the only channel that was previously measured
with any reasonable precision. The total cross section and invariant mass
distributions of the $\pi^+\pi^-$, $p\pi^+$ and $p\pi^-$ pairs were analysed in
an early attempt to extracxt the dominant production mechanism by L{\"u}ke and
S{\"o}ding \cite{Luke}. The analysis of the total cross sections and invariant
mass distributions of the pion-pion and pion-nucleon pairs clearly indicated a
dominant contribution of the $\gamma p\rightarrow\Delta^{++}\pi^-$-reaction 
via the $\Delta$-KR and the $\Delta$-pion pole terms.  
The energy dependence of the cross section thus
reflects the $\gamma N\rightarrow\pi\Delta$ threshold smeared by the width of 
the $\Delta$-resonance. More recent analysis \cite{Tejedor,Murphy},
taking into account the more precise data from the DAPHNE-detector
\cite{Braghieri} have confirmed this picture. However, it was pointed out by
Oset and coworkers \cite{Tejedor}, that although the direct contributions 
from higher resonances are negligible, the peak like structure between 
600 and 800 MeV is due to an interference of the sequential reaction
$\gamma p\rightarrow D_{13}\rightarrow\Delta\pi\rightarrow N\pi\pi$
with the leading $\Delta$-KR term.

\begin{figure}[t] 
\begin{minipage}{0.cm}
{\epsfig{file=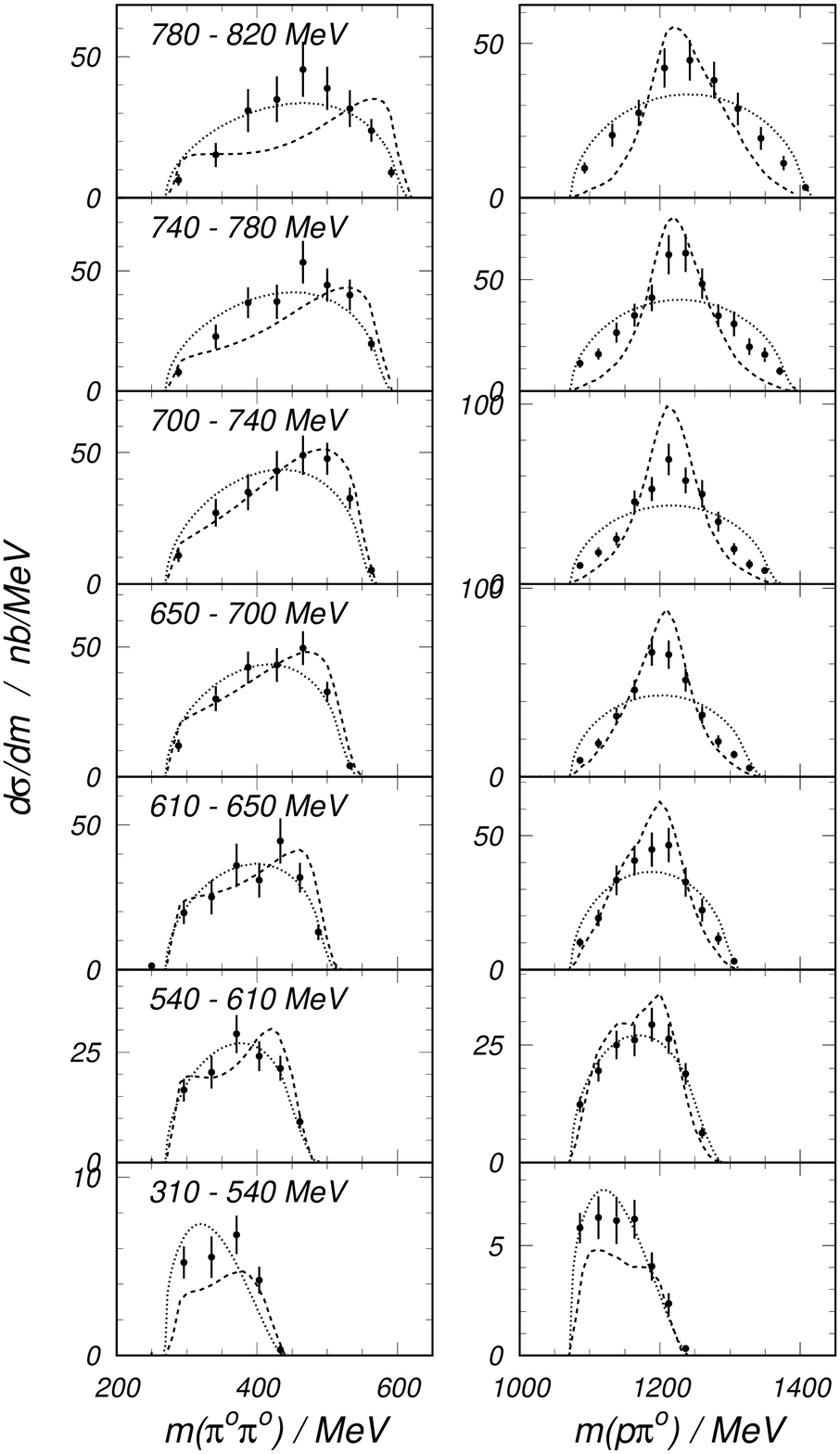,width=2.8in}}
\end{minipage}
\hspace{7.cm}
\begin{minipage}{0.cm}
{\epsfig{file=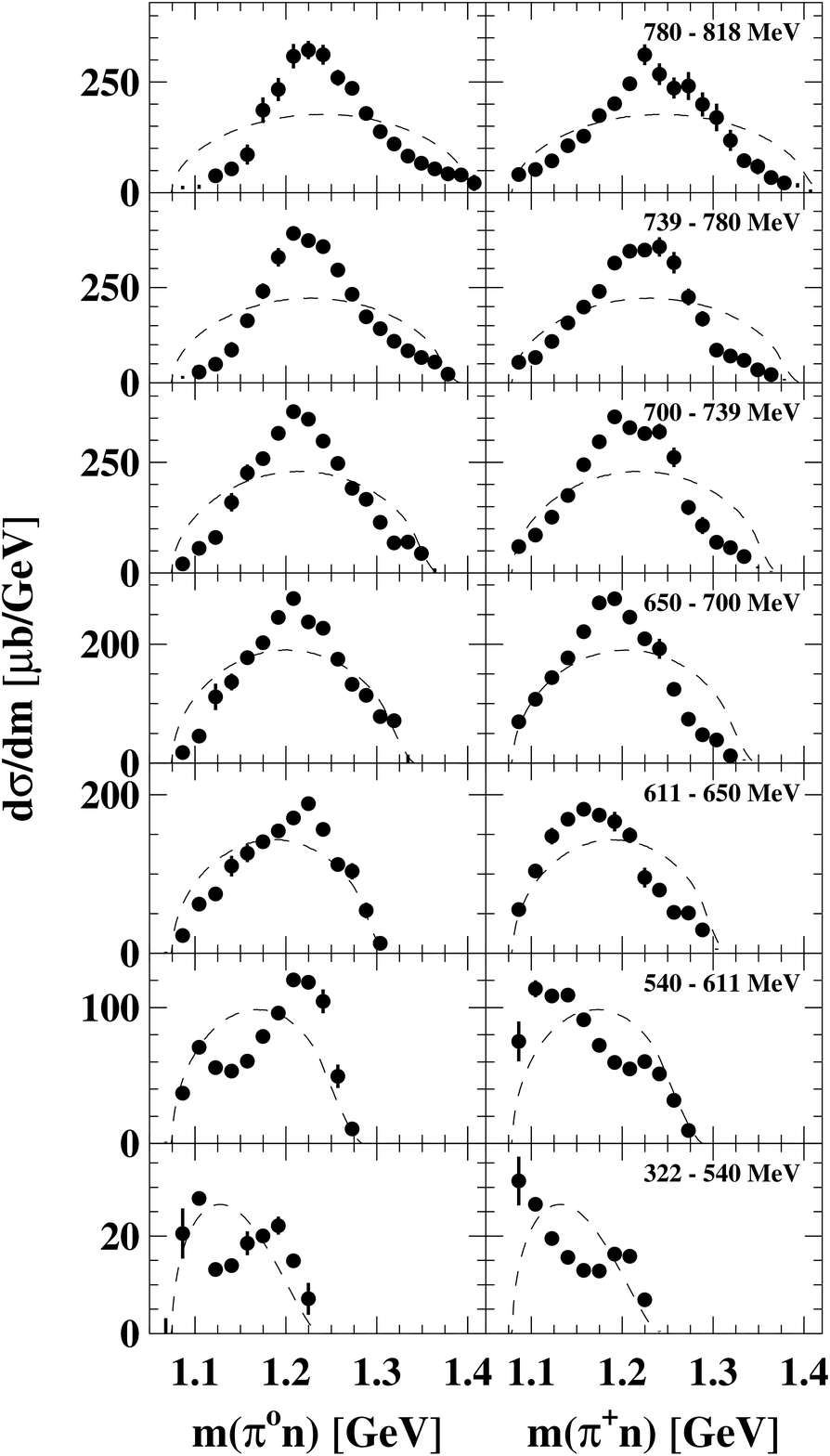,width=2.8in}}
\end{minipage}
\vspace{10pt}
\caption{Left hand side: invariant mass distributions of the pion-pion and
pion-proton pairs from the reaction 
$p(\gamma ,\pi^o\pi^o)p$. The dotted curves correspond to phase space, the
dashed curves to the model from Oset \protect\cite{Tejedor}. Right hand side: 
invariant mass distributions of the pion-neutron pairs from the 
$p(\gamma ,\pi^o\pi^+)n$ reaction. Dashed curves represent phase space behavior.
The corresponding ranges of incident photon energies are indicated in the 
pictures.  
}
\label{fig8}
\end{figure}

The situation is very different for the final states with two neutral pions.
Since the photon does not couple to neutral particles and the $\rho$-meson
does not decay into a pair of neutral pions all background terms are forbidden
or strongly surpressed. Consequently, the neutral channel is best suited for the
study of higher lying resonances. Surprisingly, the two models from refs.
\cite{Tejedor,Murphy} made very different predictions. One of them
\cite{Tejedor} predicted the sequential decay of the $D_{13}(1520)$ resonance,
the other \cite{Murphy} the decay of the $P_{11}(1440)$ resonance via a
correlated pair of pions in a relative s-wave as dominating process.
Already the total cross section \cite{Braghieri,Haerter,Wolf} is in better 
agreement with the prediction from ref. \cite{Tejedor}, but the problem was 
finally solved with the invariant mass distributions measured with TAPS 
\cite{Haerter,Wolf}. They are shown on the left hand side of Fig. \ref{fig8}.
The pion - pion invariant mass distributions follows phase space behavior, while a
strong deviation from phase space behavior was predicted for the correlated
two pion decay of the $P_{11}$ in \cite{Murphy}. On the other hand the
pion - proton invariant mass deviates from phase space and peaks at the $\Delta$
mass as expected for a sequential 
$N^{\star}\rightarrow\Delta\pi^o\rightarrow\ N\pi^o\pi^o$ decay and as
predicted in \cite{Tejedor}.  
Since the two neutral pions are indistinguishable, even in case of a pure 
sequential resonance decay the spectra are composed of 
the $\Delta$-component and a phase space like component from the combination 
of the 'wrong' $\pi^o$ with the proton. 
The high quality invariant mass distributions
available now, will certainbly allow a more detailed analysis, e.g. the
predictions from \cite{Tejedor} start to deviate from the measured $\pi^o\pi^o$
invariant masses at photon energies above 750 MeV.

The situation was even more puzzling for the $\pi^o\pi^+$-channel.
The first measurement of the $\gamma p\rightarrow\pi^o\pi^+$ reaction 
\cite{Braghieri} came up with a total cross section that was strongly 
underestimated by the predictions from the then available models
\cite{Tejedor,Murphy} (see figure \ref{fig7}, right hand side).
In the meantime this finding was confirmed by a measurement with the TAPS
detector \cite{Langgaertner} and a similar situation was found for the
$\gamma n\rightarrow p\pi^-\pi^o$ reaction \cite{Zabrodin}. Obviously an 
important contribution is severely underestimated in the models.

\begin{figure}[t] 
\begin{minipage}{0.cm}
{\epsfig{file=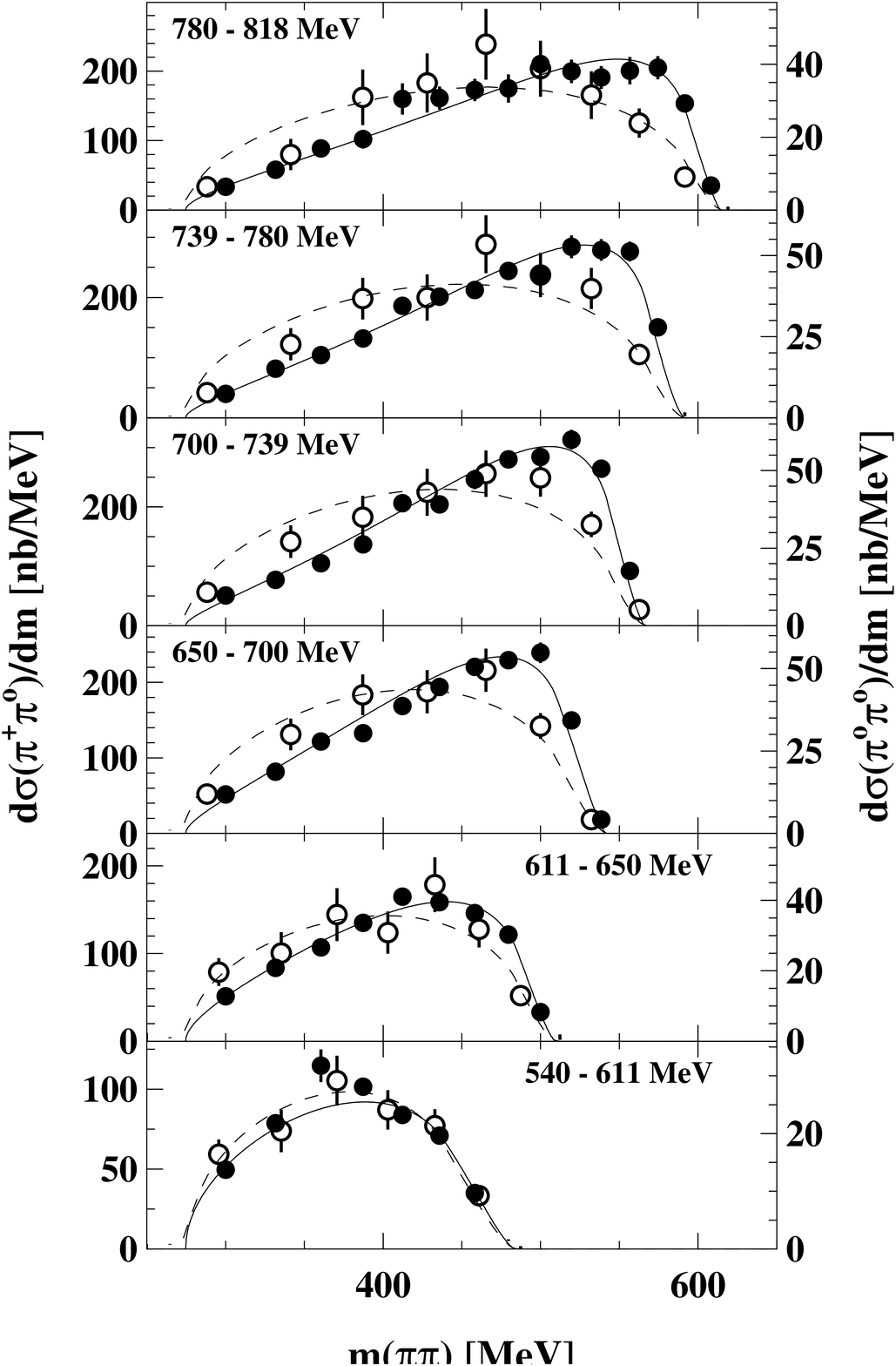,height=4.1in,width=2.8in}}
\end{minipage}
\hspace{7.5cm}
\begin{minipage}{0.cm}
{\epsfig{file=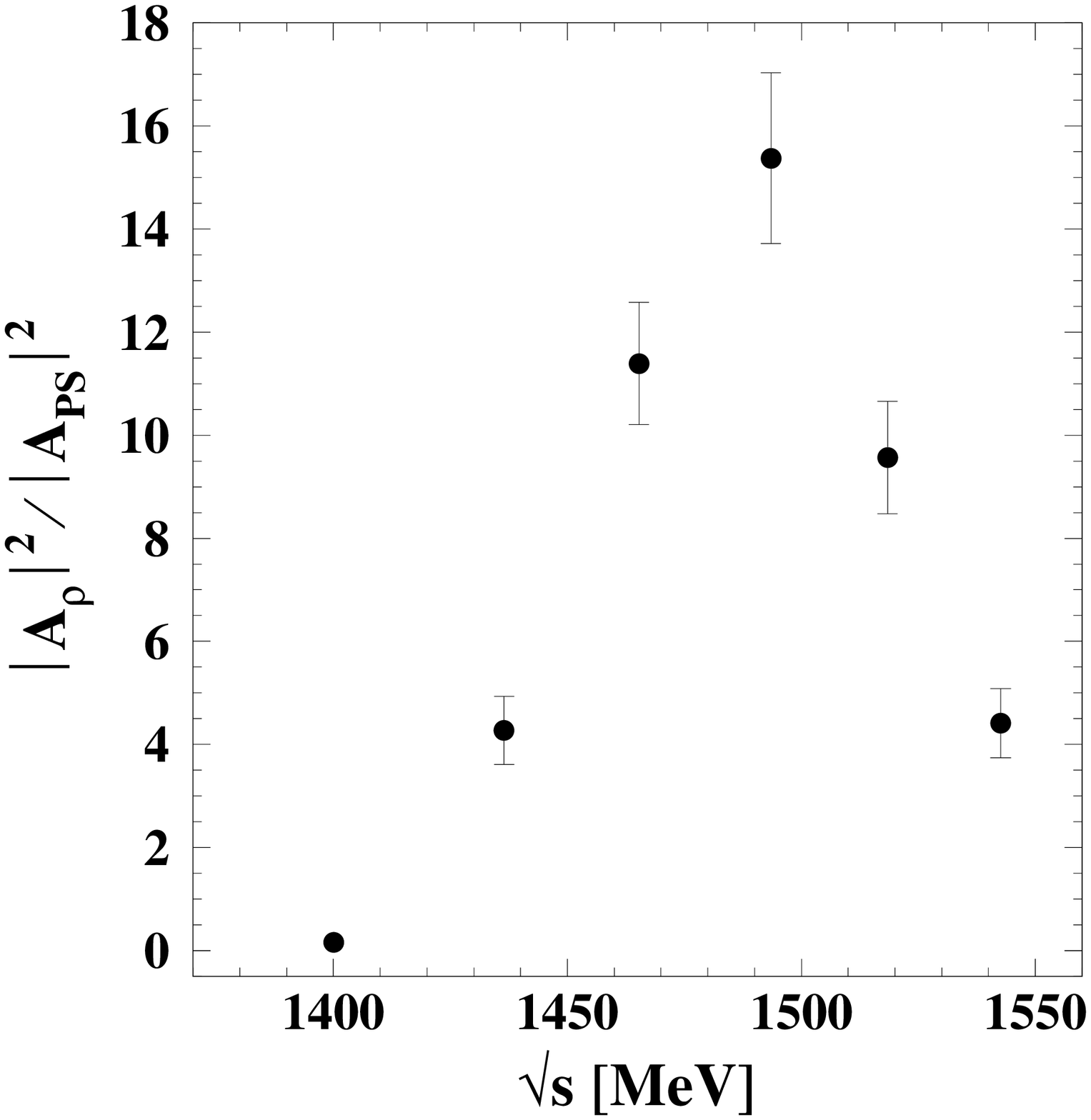,height=4.1in,width=2.8in}}
\end{minipage}
\vspace{10pt}
\caption{Left hand side: invariant mass distributions of the pion-pion 
pairs from the reactions $p(\gamma ,\pi^o\pi^o)p$ (open circles, scale at right
axis) and $p(\gamma ,\pi^o\pi^+)n$ (filled circles, scale at left axis). The dashed
curves curves correspond to phase space, the full curves to the fit with phase
space and $N\rho$ components (see text). The picture at the right hand side
shows the ratio of the matrix elements for $N\rho$- to phase space components 
obtained from the fit. 
}
\label{fig9}
\end{figure}
 
Recently Ochi et al. \cite{Ochi} suggested that a large contribution of 
the $\rho$-Kroll-Rudermann term which is negligible for the other isospin 
channels might solve the problem. However, in many other aspects their model 
is more
simplifying than the others and describes the other charge channels less well.
Nevertheless, this suggestion motivated a carefull study of the invariant
mass distributions of the pion - pion and the pion - nucleon pairs from this
reaction, which are again the most sensitive observables. The invariant mass
distributions of the two possible pion - neutron combinations are shown at the
right hand side of Fig. \ref{fig8}. The $\pi^o n$ invariant mass peaks already
at low incident photon energies at the $\Delta$ mass, but this signal does not
appear in the $\pi^+ n$ invariant mass at low photon energies. This behavior is
expected if, as predicted in the models, the process is dominated at low 
photon energies by the $\Delta$-KR- and $\Delta$-pion-pole terms. Since the
photon does not couple to the neutral pion, the charged pion is produced at the
first vertex and the neutral pion stems from the subsequent 
$\Delta^o\rightarrow N\pi^o$ decay, giving rise to the $\pi^o n$ invariant mass
correlation. At higher incident photon energies  sequential
decays of $N^{\star}$ resonances may contribute like in $\pi^o\pi^o$.
It is easely seen from the relevant Clebsch-Gordan coefficients, that in this
case both sequences of the charged and neutral pion are equally probable, so
that the $\Delta$-signal may also appear in the $\pi^+ n$ invariant mass.

Possible contributions from $\rho^+$-meson production should be visible in the
pion - pion invariant mass distribution as an enhancement towards high invariant
masses. The DAPHNE collaboration \cite{Zabrodin} has searched for such 
enhancements in the quasifree $d(\gamma, \pi^o\pi^-)pp$ reaction and indeed
found some indication of the effect. However, their analysis is largely
complicated by effects from the bound nucleons. The pion - pion invariant mass
distributions for the $\pi^o\pi^o$ and $\pi^o\pi^+$ pairs from the free proton
measured with TAPS \cite{Wolf,Langgaertner} are compared at the left hand side
of Fig. \ref{fig9}. The comparison of these two channels is particulary
instructive since the $\rho^o\rightarrow\pi^o\pi^o$ decay is forbidden so that
the $\rho$-meson cannot contribute to the double $\pi^o$ channel. As already
mentioned, the $\pi^o\pi^o$ invariant mass shows phase space behavior, but at
the higher incident photon energies the $\pi^o\pi^+$ invariant mass is clearly 
shifted to large masses as expected for off-shell $\rho^+$ contributions.
The $\pi^o\pi^+$-data were fitted with a simple model assuming only phase space
and $\rho$-decay contributions \cite{Langgaertner} via:
\begin{equation}
\frac{d\sigma}{dm}\propto
|a(\sqrt{s})+b(\sqrt{s})p_{\pi}(m_{\pi\pi})D_{\rho}(m_{\pi\pi})|^2
ps_{\sqrt{s}\rightarrow\pi\pi N}
\end{equation}
where $p_{\pi}$ is the momentum of the $\pi$ in the $\rho$ rest frame, 
$ps_{\sqrt{s}\rightarrow\pi\pi N}$ is the three body phase space factor and 
$D_{\rho}$ represents the $\rho$-meson propagator. The constants $a$ and $b$
where fitted to the data and the ratio of the matrix elements for $\rho$-meson
decays and phase space components was calculated via:
\begin{equation}
\frac{|A_{\rho}|^2}{|A_{ps}|^2}=
\frac{\int|b(\sqrt{s}) p_{\pi}(m_{\pi\pi})D_{\rho}(m_{\pi\pi})|^2dm_{\pi\pi}}
{\int|a(\sqrt{s})|^2dm_{\pi\pi}}
\end{equation}
The result of the ratio of the matrix elements is shown at the right hand side
of Fig. \ref{fig9}. Note that this is the ratio of the matrix elements without
phase space factors. 
The relative contribution of the $\rho$-decay matrix element peaks close to the
position of the $D_{13}$ resonance which may hint at a significant
$D_{13}\rightarrow N\rho$ contribution to $\pi^o\pi^+$-photoproduction.
This component was until now omitted in the model calculations. New calculations
of the $\pi^o\pi^+$-photoproduction reaction in the framework of \cite{Tejedor}
including this component are under way and the comparison to the present data
should allow a more precise extraction of the $D_{13}\rightarrow N\rho$
branching ratio (PDG estimate: 15-25\% \cite{PDG}).

\subsubsection{The second resonance region for bound nucleons}  
The problem of the suppression of the second resonance bump in total
photoabsorption on nuclei was already mentioned. By now it should be
clear, that the complicated structure of this 'bump' makes a detailed
investigation of exclusive meson decay channels highly desirable.

Some years ago,
we have used $\eta$-photoproduction for an investigation of the 
in-medium properties of the $S_{11}(1535)$ \cite{Roebig}, but did not find any 
unexplained in-medium effects. The data were in 
excellent agreement with  model predictions taking into account
trivial nuclear effects. Since the energy range extended just up
to the resonance maximum, it was not possible to deduce the in-medium width of
the resonance. In the meantime Yorita et al. \cite{Yorita} have studied
this reaction from carbon over a larger energy range and
found no clear broadening of the $S_{11}$ resonance, again the data are 
in fair agreement with model expectations.

However, the total contribution of the $S_{11}$-resonance to the
bump structure is quite small. Furthermore due to the location of the 
$\eta$-production threshold on the resonance, nuclear Fermi motion 
has a large influence on the excitation curve so that any extraction of
the in-medium width requires a lot of modelling. 

Since most of the tentative explanations of the disappearing of the resonance
bump involve a significant broadening of the $D_{13}$ resonance, we have now 
investigated the $D_{13}(1520)$ resonance via quasifree 
single $\pi^o$-photoproduction from nuclei. The neutral pions were identified 
via an invariant mass analysis and 
quasifree single $\pi^o$-production was selected by a missing energy analysis
as in \cite{Krusche_4}. The stronger broadening (compared to the deuteron target
\cite{Krusche_4}) of the structures in the missing energy spectra by Fermi 
motion was compensated by more restrictive cuts.

\begin{figure}[h] 
\begin{minipage}{0.cm}
{\epsfig{file=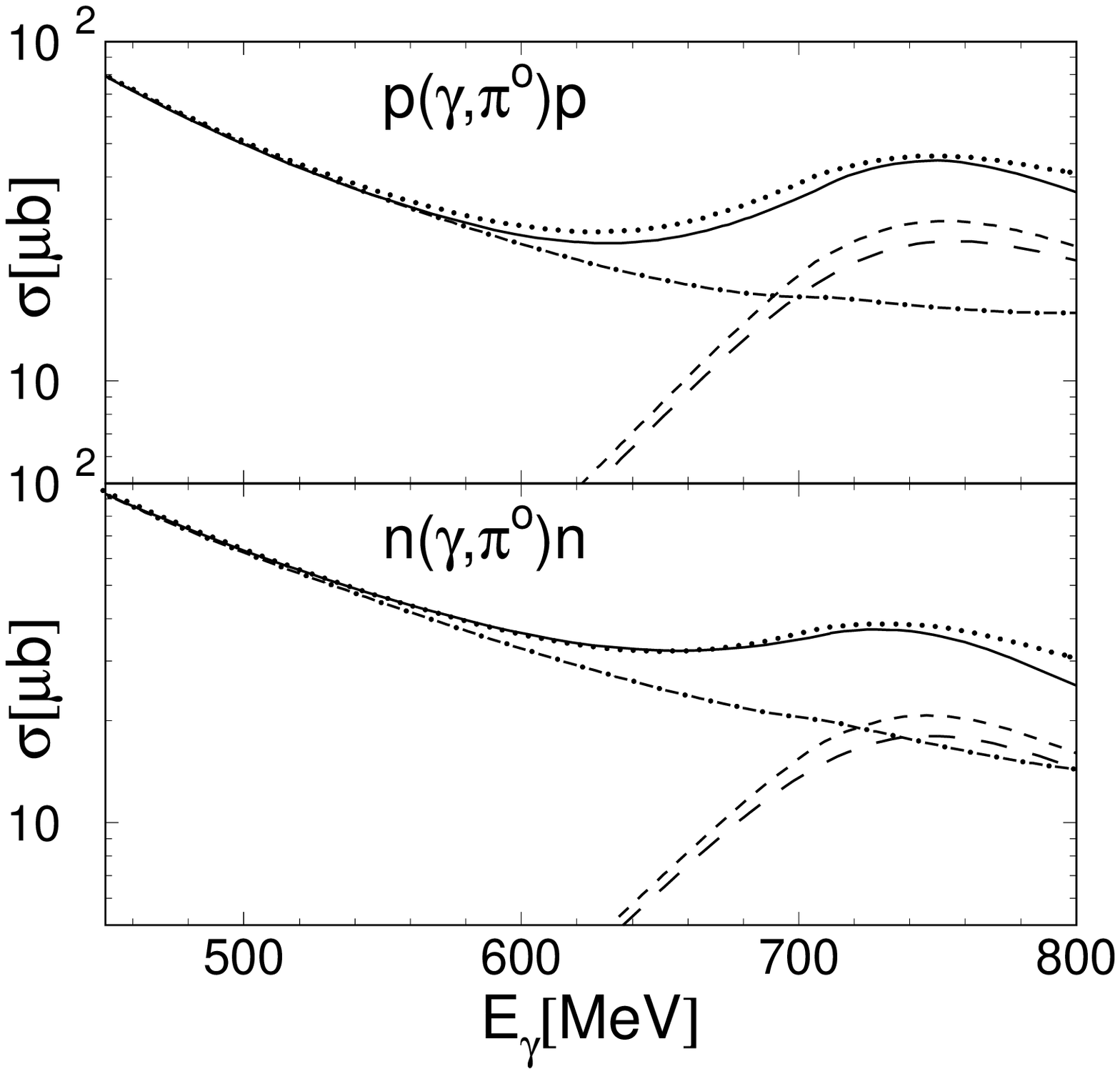,width=2.8in}}
\end{minipage}
\hspace{7.cm}
\begin{minipage}{0.cm}
{\epsfig{file=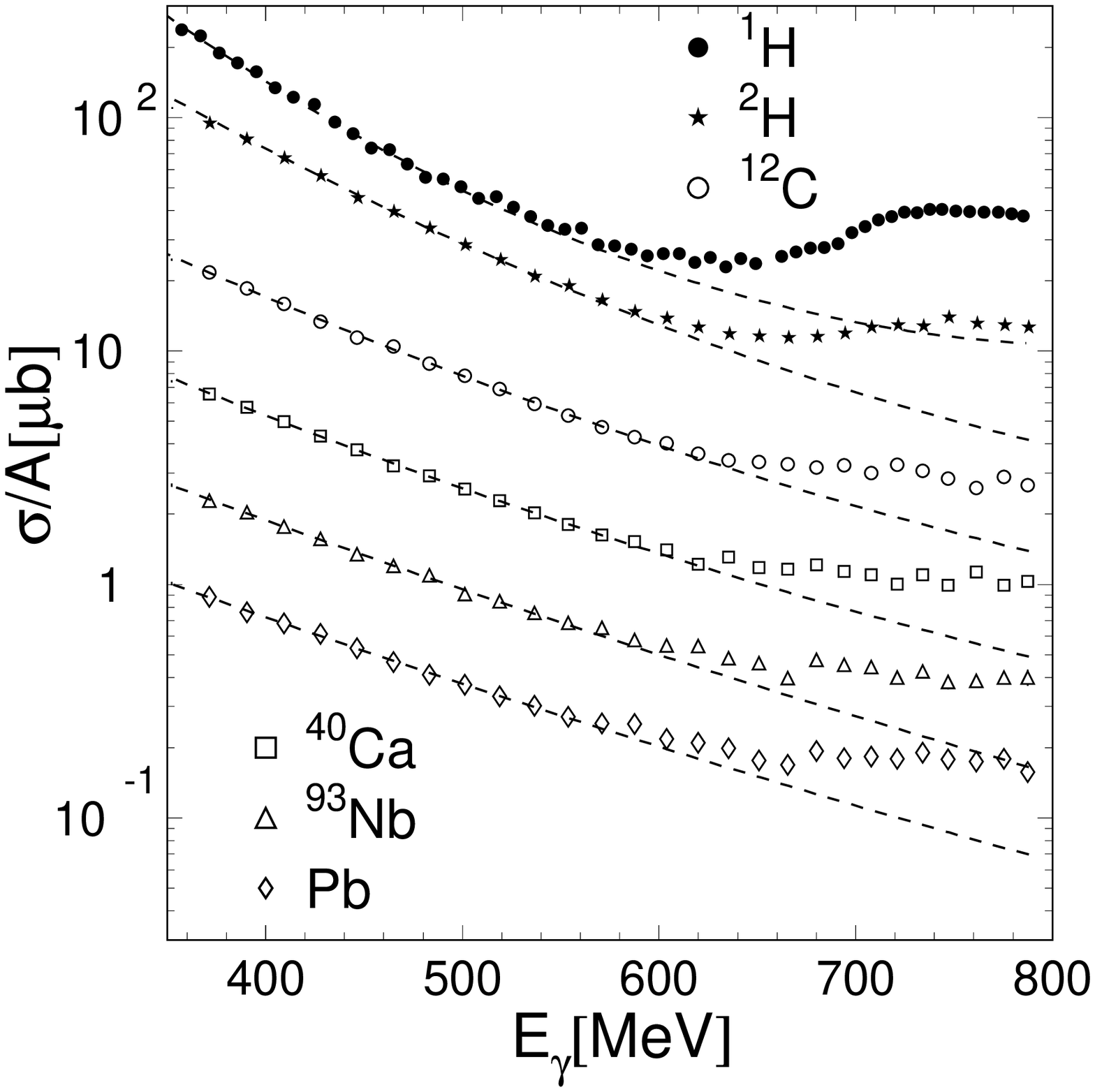,width=2.8in}}
\end{minipage}
\vspace{10pt}
\caption{Left hand side:
Decomposition of single $\pi^o$ photoproduction from MAID2000 
\protect\cite{MAID}. Full lines: cross section $\sigma_{\pi^o}$ for full 
model; dash-dotted curves: cross section $\sigma_{nr}$ without contribution
from $D_{13}$ and $S_{11}$, short-dashed curves: cross section
$\sigma_{r}$ for excitation of $D_{13}$ and $S_{11}$ only, long-dashed curves:
$D_{13}$ only, dotted curves: $\sigma_{nr}+\sigma_{r}$.
Right hand side: Preliminary total cross section per nucleon 
for single $\pi^o$ production for the 
nucleon and for nuclei. Scale corresponds to the proton data, other 
data scaled down by factors 2,4,8,16,32. Dashed curves: fits to the
data in the energy range 350 - 550 MeV.   
}
\label{fig10}
\end{figure}

We have
decomposed the cross sections into a resonance and a background part. In
principle such a decomposition requires a multipole analysis wich takes into
account resonance - background interference terms. However, as demonstrated in
the left hand side of Fig. \ref{fig10} interference terms are small in this 
case. Shown are for the proton and the neutron the cross sections calculated  
from a unitary isobar analysis of pion photoproduction (MAID) \cite{MAID} 
taking into account all resonances and background terms ($\sigma_{\pi^o}$).
They agree quite well with the sum of the cross sections $\sigma_r$ 
(excitation of the $S_{11}$ and $D_{13}$ resonances only) and $\sigma_{nr}$
(everything except $S_{13}$ and $D_{13}$ excitation) also obtained with MAID.

The decomposition of the measured cross sections is shown in the right hand 
side of Fig. \ref{fig10}. The background part coming from the tail of the
$\Delta$-resonance, the contribution of the $P_{11}$-resonance, nucleon Born
terms and vector meson exchange was fitted with a function of the type
\begin{equation}
\sigma\propto exp(aE_{\gamma}^2+bE_{\gamma}). 
\end{equation}
with $a$ and $b$ as free parameters. 
The quantitative analysis of the data is still in progress, however the results
obtained so far for the nuclear targets do not exhibit any obvious broadening 
in addition to Fermi smearing compared to the reaction from the deuteron. 

\section*{Summary} 
The properties of nucleon resonances excited on the free nucleon or inside
nuclear matter are one of the most important testing grounds for models of the
strong interaction in the non-perturbative regime. The present talk has
concentrated on new, precise measurements of properties of the low 
lying nucleon resonances, in particular the $\1r$-, $\3r$- and $\4r$-resonances
via meson production experiments. Experiments on the free nucleon like e.g.
the measurements of the $E2$-admixture of the $\Delta$-excitation, the magnetic 
moment of the $\Delta$, the branching ratios of the $D_{13}$-resonance or the
neutron/proton ratio of the $S_{11}$ helicity amplitudes are
approaching a new level of sensitivity and precision. The in-medium properties
of the resonances are still not understood. The total photoabsortpion
experiments show an almost complete depletion of the second resonance bump
but exclusive meson production experiments show no indication for a significant
broadening of the $S_{11}$ or $D_{13}$ resonances.


\begin{references}
\bibitem{Krusche_1}B. Krusche et al., {\it Phys. Rev. Lett.}\ {\bf 74}, 
3736 (1995).
\bibitem{Ajaka}J. Ajaka et al., {\it Phys. Rev. Lett.} {\bf 81}, 1797 (1998).
\bibitem{Tiator} L. Tiator et al., {\it Phys. Rev.} {\bf C60} 35210 (1999).
\bibitem{PDG} C. Caso et al., {\it Eur. Phys. J.} {\bf C3} 1 (1998).
\bibitem{Langgaertner} W. Langg\"artner it et al., {\it submitted to Phys. Rev.
Lett.}
\bibitem{Frommhold} Th. Frommhold et al., {\it Phys. Lett.} {\bf B295} 28
(1992); {\it Z. Phys.} {\bf A350} 249 (1994).
\bibitem{Bianchi} N. Bianchi et al., {\it Phys. Lett.} {\bf B299} 219 (1993).
\bibitem{Beck97} R. Beck et al., {\it Phys. Rev. Lett.}, {bf 78} 606 (1997).
\bibitem{Blanpied97} G. Blanpied et al., {\it Phys. Rev. Lett.}, {\bf 79} 4337
(1997).
\bibitem{Beck00} R.Beck et al., {\it Phys. Rev.}, {\bf C61} 35204 (2000).
\bibitem{Nefkens} B.M.K. Nefkens et al., {\it Phys. Rev.} {\bf D18} 3911 (1978).
\bibitem{Bosshard} A. Bosshard et al., {\it Phys. Rev.} {\bf D44} 1962 (1991).
\bibitem{Machavariani} A.I. Machavariani et al., {\it Nucl. Phys. A646} 231 
(1999) and erratum (numerical results must be devided by factors $(4\pi )^2$,
$(4\pi )^3$).
\bibitem{Drechsel0} D. Drechsel et al., {\it Phys. Lett.} {\bf B484} 236 (2000).
\bibitem{Kotulla} M. Kotulla, U. Giessen, priv. com.
\bibitem{Krusche_2} B. Krusche et al., {\it submitted to PLB}.
\bibitem{Drechsel} D. Drechsel  et al., Nucl. Phys. {\bf A660} 423 (1999), 
         and priv. com. S. Kamalov.
\bibitem{Drechsel2} D. Drechsel  et al., Nucl. Phys. {\bf A645} 145 (1999). 
\bibitem{Rambo}F. Rambo  et al., Nucl. Phys. {\bf A660} 69 (2000).
\bibitem{Krusche_2} B. Krusche it et al., Phys. Lett. {\bf B397} 171 (1997).
\bibitem{Bock} A. Bock et al., Phys. Rev. Lett. {\bf 81} 534 (1998).
\bibitem{Krusche_3} B. Krusche et al., Phys. Lett. {\bf B358} 40 (1995).
\bibitem{Hoffmann} P. Hoffmann-Rothe et al., Phys. Rev. Lett. {\bf 78} 4697
(1997).
\bibitem{Hejny} V. Hejny et al., Eur. Phys. J. {\bf A6} 83 (1999).
\bibitem{Braghieri} A. Braghieri et al., Phys. Lett. {\bf B363} 46 (1995).
\bibitem{Tejedor} J.A. Gomez Tejedor, et al., Nucl. Phys. {\bf A600} 413
(1996).
\bibitem{Haerter} F. Haerter et al., Phys. Lett. {\bf B401} 229 (1997).
\bibitem{Zabrodin} A. Zabrodin et al., Phys. Rev. {\bf C55} R1617 (1997);
                   Phys. Rev {\bf C60} 5201 (1999).
\bibitem{Krusche_4} B. Krusche et al., Eur. Phys. J. {\bf A6} 309 (1999).
\bibitem{Kleber} V. Kleber et al., Eur. Phys. J. {\bf A9} 1 (2000).
\bibitem{Wolf} M. Wolf et al., Eur. Phys. J. {\bf A9} 5 (2000).
\bibitem{Langgaertner} W. Langg\"artner et al., submitted to Phys. Rev.
Lett.
\bibitem{Wilhelm} B.Schoch et al., {\it Prog. Part. Nucl. Phys.} {\bf 34} 43 (1995).
\bibitem{Sauermann}C. Deutsch-Sauermann et al., {\it Phys. Lett.} {\bf 409}
1 (1997).
\bibitem{Dytmann} S.A. Dytmann et al., {\it Proc. IV CEBAF/INT Workshop
on $N^*$-physics, World Scientific, eds. T.-S.H. Leee and W. Roberts}, p286
(1997).
\bibitem{Yorita} T. Yorita et al., {\it Phys. Lett.} {\bf B 476} 226 (2000).
\bibitem{Weiss} J. Weiss, PhD thesis, University Giessen 1999, to be published.
\bibitem{Murphy} L.Y. Murphy et al., {\it DAPHNIA/SPhN}, {\bf 96-10} 1 1996.
\bibitem{Ochi} K. Ochi et al., {\it Phys. Rev.}, {\bf C56} 1472 1997.
\bibitem{Giannini} M.M. Giannini et al., {\it Phys. Rev.} {\bf C49} R1258
(1994).
\bibitem{Lehr} J. Lehr  et al. Phys. {\bf A671} 503 2000.
\bibitem{Luke} D.~L{\"uke} et al.,{\it Springer Tracts in Modern Physics} 
{\bf 59} 39 (1971).
\bibitem{Roebig} M. R\"obig-Landau et al., Phys. Lett. {\bf B373} 45 (1996).
\bibitem{MAID} D. Drechsel et al., Nucl Phys. {\bf A645} 145 (1999).
http://www.kph.uni-mainz.de/MAID/maid.html. 
\end{references}
\end{document}